\documentclass[natbib]{svjour3}            
\smartqed  
\usepackage{graphicx}
\usepackage{aps-bibstyle}  
\newcommand{\la}{\raise0.3ex\hbox{$<$}\kern-0.75em{\lower0.65ex\hbox{$\sim$}}}
\newcommand{\ga}{\raise0.3ex\hbox{$>$}\kern-0.75em{\lower0.65ex\hbox{$\sim$}}}
\journalname{Space Science Reviews}
\newcommand{\apj}{{Astrophys. J.}\ }
\newcommand{\apjl}{{Astrophys. J. Lett.}\ }
\newcommand{\aap}{{Astron. Astrophys.}\ }
\newcommand{\mnras}{{Mon. Not. Royal Astron. Soc.}\ }

\newcommand{\araa}{{Ann. Rev. Astron. Astrophys.}\ }
\newcommand{\prl}{{Phys. Rev. Lett.}\ }
\newcommand{\prd}{{Phys. Rev. D}\ }
\newcommand{\pre}{{Phys. Rev. e}\ }
\newcommand{\physrep}{{Phys. Rep.}\ }
\begin{document}

\title{Magnetic fields in the Large-Scale Structure of the Universe}

\titlerunning{Magnetic fields in LSS} 

\author{D. Ryu \and D. R. G. Schleicher \and
R. A. Treumann \and C. G. Tsagas \and L. M. Widrow}

\authorrunning{Ryu et al.} 

\institute{Dongsu Ryu \at
Department of Astronomy and Space Science,
Chungnam National University, Daejeon 305-764, Korea,
\email{ryu@canopus.cnu.ac.kr}
\and
Dominik R. G. Schleicher \at
Georg-August-Universit\"at,
Institut f\"ur Astrophysik,
Friedrich-Hund-Platz 1,
37077 G\"ottingen,
Germany,
\email{dschleic@astro.physik.uni-goettingen.de}
\and
Rudolf A. Treumann \at
ISSI, CH-3012 Bern, Hallerstrasse 6, Switzerland,
\email{treumann@issibern.ch}
\and
Christos G. Tsagas \at Department of Physics,
Aristotle University of Thessaloniki, Thessaloniki 54124, Greece,
\email{tsagas@astro.auth.gr}
\and
Lawrence M. Widrow \at
Department of Physics, Queen's University, Kingston, Ontario, K7L 3N6, Canada,
\email{widrow@astro.queensu.ca}}

\date{Received: date / Accepted: date}

\maketitle

\begin{abstract}

Magnetic fields appear to be ubiquitous in astrophysical
environments.  Their existence in the intracluster medium is
established through observations of synchrotron emission and Faraday
rotation.  On the other hand, the nature of magnetic fields outside
of clusters, where observations are scarce and controversial,
remains largely unknown.  In this chapter, we review recent
developments in our understanding of the nature and origin of
intergalactic magnetic fields, and in particular, intercluster
fields.  A plausible scenario for the origin of galactic and
intergalactic magnetic fields is for seed fields, created in the
early universe, to be amplified by turbulent flows induced during
the formation of the large scale structure.  We present several
mechanisms for the generation of seed fields both before and after
recombination.  We then discuss the evolution and role of magnetic
fields during the formation of the first starts.  We describe the
turbulent amplification of seed fields during the formation of large
scale structure and the nature of the magnetic fields that arise.
Finally, we discuss implications of intergalactic magnetic fields.


\end{abstract}

\section{Introduction}\label{s-intr}

In the highly successful $\Lambda$CDM cosmology, the large scale structure
(LSS) forms through a process known as hierarchical clustering in which
small-scale objects collapse first and merge to form systems of
ever-increasing size.  This scenario leads to a ``cosmic web'' of
structure where galaxies reside mainly along filaments while galaxy
clusters arise at the intersections of two or more filaments
\citep[see, e.g.,][]{bkp96}.  Furthermore, a picture of a multi-phase
intergalactic medium (IGM) has emerged.  A hot phase, often referred
to as the intracluster medium (ICM) because it is found inside and
around clusters and groups of galaxies, has a temperature $T > 10^7$ K
and is observable via X-ray emission.  The warm-hot intergalactic
medium (WHIM) ($10^5 < T < 10^7$ K) is found mainly in filaments of
galaxies \citep{co99,krcs05}.  Gas in the IGM is heated by
cosmological shocks which arise as objects in the hierarchy form.
\citep{rkhj03,psej06,krco07,sohb08,vbg09}.

Magnetic fields are observed in galaxies of all types and in galaxy
clusters.  Moreover, there is some evidence that they permeate the
filaments of the cosmic web.  Various scenarios for the origin of
galactic magnetic fields are discussed in \cite{wrss10} of this volume.
The basic
idea is that weak fields, generated either by an exotic early universe
mechanism or some astrophysical process, are amplified to $\mu$G
strength during galaxy formation and by dynamo action during the
subsequent quiescent phase of galaxy evolution.  Similarly, weak seed
magnetic fields can be amplified into the intergalactic magnetic field
(IGMF) by turbulent flow motions during the formation of the LSS
through a process known as small-scale turbulence dynamo.  In a
turbulence dynamo, kinetic energy of the fluid in converted to
magnetic energy through stretching, twisting, and folding of the field
\citep[see, e.g.,][of this volume]{Subramanian99,cv00,Haugen04a,Haugen04b,
Haugen04c,Schekochihin04,Brandenburg05,cvbl09,bss10}.
In addition, feedback from the black hole regions in active galactic
nuclei (AGNs) can also contribute and amplify magnetic fields in the IGM.

This chapter will explore the structure formation-magnetic field
connection.  In particular, we will address two interrelated
questions.  First, can the structure formation itself generate new
magnetic fields and amplify existing ones?  Second, what role do
magnetic fields play in structure formation and in other astrophysical
processes?  An outline of the chapter is as follows: In Section 2, we
briefly review the observational evidence for magnetic fields in the
clusters and filaments of the cosmic web.  In Section 3, we describe
various astrophysical mechanisms for the generation of seed magnetic
fields, which would later be maintained and amplified within galaxies
and clusters.  In Section 4, we discuss the evolution and role of
magnetic fields during the first star formation: their stabilization
due to a non-zero ionization degree as well as their potential
implication for star formation.  We then describe how magnetic fields
are amplified during structure formation in Section 5. We discuss the
effects of magnetic fields on the propagation of cosmic rays in the
IGM and also the Faraday rotation measure (RM) induced by magnetic
fields in Section 6.  Section 7 presents something of a departure from
the more phenomenological and astrophysical discussion as it provides
a formal treatment of the evolution of density perturbations in the
presence of magnetic fields.  Finally, a brief summary follows in
Section 8.

\section{Observational Evidence for Magnetic Fields in Clusters and
Filaments}\label{s-obs}

The existence and strength of the IGMF has been deduced from
observations of synchrotron emission and confirmed directly from
observations of RM \citep[see, e.g.,][for review]{ct02,gf04}.
Synchrotron emission from a galaxy
cluster was first discovered by \citet{lmh59} who surveyed the Coma
cluster in the radio.  Since then, it has been observed in numerous
clusters, either as radio halos or as radio relics \citep[see,
e.g.,][and references therein]{gf04,cbvs08}.  The diffuse radiation
from radio halos is mostly unpolarized, and is thought to emerge from
turbulent magnetic fields in the ICM.  The radiation from radio relics
is highly polarized and is believed to be emitted from shocked
regions in the ICM.  The strength of the magnetic field in radio
halos is estimated to be $\sim 1 \mu$G, while stronger fields are found
radio relics.  These estimates either assume equipartition (also known
as the minimum energy argument) or incorporate measurements of
inverse-Compton emission.

Observations of the IGMF based on Faraday rotation have also been done,
though mostly for magnetic fields in the ICM \citep[see][and
references therein]{ct02}.  An RM map of the Coma cluster, for
example, reveals a field with a strength $\sim\mu$G and a coherence
length of order $\sim 10$\ kpc \citep{kkdl90}.  For Abell clusters,
the typical RM is $\sim100 -200\ {\rm rad\ m^{-2}}$ which indicates an
average field strength of $\sim 5$--10\ $\mu$G \citep{ckb01,cla04}.
RM maps of clusters can be used to determine the power spectrum of
turbulent magnetic fields in the ICM.  For instance, a Kolmogorov-like
spectrum with a bending at a few kpc scale is found in the cooled
core region of the Hydra cluster \citep{ve05}, and spectra consistent
with the Kolmogorov spectrum were reported in the wider ICM for the
Abell 2382 cluster \citep{gmgp08} and for the Coma cluster
\citep{bfmg10}.

The nature of the IGMF in filaments, on the contrary, remains largely
unknown, because the studies of synchrotron emission and RM outside
clusters are still scarce and controversial.  Although faint radio
emission has been observed in the outskirts of clusters \citep[see,
e.g.,][]{kkgt89}, there are no confirmed observations of synchrotron
emission from filaments.  Such measurements present a challenge for
current facilities.  As well, the removal of the galactic foreground
is a non-trivial task \citep[see, e.g.,][]{bfr10}.  At present there
is an upper limit of $\sim 0.1\ \mu$G for the strength of the IGMF in
filaments based on the observed limit of the RMs of background quasars
\citep{rkb98,xkhd06}.

Recently, \citet{nv10,alek10} reported a lower bound for the strength of
the magnetic field in voids.  Their claim, if true, is significant as it
would represent the first evidence for magnetic fields in such low-density
regions and on such large scales.  The basic idea is that an IGMF will
deflect charged particles that arise in an electromagnetic cascade whose
source is the very high energy $\gamma$-rays produced in an AGN.  A
non-observation of $\gamma$-ray secondaries coincident with VHE
$\gamma$-rays, is then taken as evidence of magnetic deflection along the
path from the source to the observer.  \citet{nv10} quote a lower bound of
$B\ge 3\times 10^{-16}\,{\rm G}$ based on observations by the Fermi and
the High Energy Stereoscopic System (HESS) telescopes.  A similar analysis
was performed by \citet{alek10} using data from the MAGIC telescope though
they are more cautious in presenting their conclusions and model
dependencies.
Note that \citet{nv10} assume continuous emission of gamma-rays for $10^6$
years, or longer which may not be realistic.  If this assumption is
relaxed, then one obtains a more conservative lower bound \citep{dermer}.

The detection of the IGMF in filaments, if it exists, might be made
with the next generation radio facilities.  These facilities include
the Square Kilometer Array (SKA), and upcoming SKA pathfinders, the
Australian SKA Pathfinder (ASKAP) and the South African Karoo Array
Telescope (MeerKAT), as well the Low Frequency Array (LOFAR)
\citep[see, e.g., papers in][]{cr04}.

For discussions on magnetic fields in the ICM and cluster outskirts,
see \citet{sbfk10} and \citet{bbrp10} of this volume.

\section{Plasma Physics Mechanisms for Seed Fields}\label{s-plas}

\subsection{Biermann battery}\label{s-batt}

The Biermann battery is a promising mechanism for the creation of
astrophysical magnetic fields.  The mechanism arises in an ionized
plasma whenever baroclinity exists, that is, when isodensity surfaces
do not coincide with isobaric surfaces.  This situation leads to an
extra pressure gradient term in Ohm's law which drives currents.
These currents, in turn,
generate magnetic fields at a rate given by
\begin{equation}
\frac{d\vec B}{dt} \sim \frac{m_e c}{e}
\frac{{\vec\nabla}\rho_e\times{\vec\nabla}p_e}{\rho_e^2}~,
\label{bier-e01}
\end{equation}
where $\rho_e$ and $p_e$ are the electron density and pressure,
respectively \citep{bier50}.
Though the mechanism was originally studied in the context of stars
\citep{bier50}, it may arise during structure formation in cosmology
whenever the electron pressure and density gradients are not aligned, as
often occurs in shocks \citep{Pudritz89,kcor97,dw00,xocn08}.

We note that vorticity, $\vec\omega$, is generated when the total pressure
and density gradients are not aligned (baroclinity of flows -- see
Equation \ref{ryu-e02} below).  In an ionized plasma,
a simple order of
magnitude estimate yields
\begin{equation}
B \sim \frac{m_p c}{e}\omega\ \simeq\ 3\times 10^{-21}\
\left (\frac{\omega}{\rm km \,s^{-1}\ kpc^{-1}}\right ) \ {\rm Gauss}~.
\label{bier-e02}
\end{equation}
Since the present-day vorticity in the IGM is of order $\sim$ a few km
s$^{-1}$kpc$^{-1}$ (see Section \ref{s-turb} below), the seed fields
from the Biermann battery will be rather small ($\sim 10^{-20}$ G).
However, vorticity, and hence seed fields, were almost certainly
larger at early times.  The following argument, based on dimensional
analysis, suggests that for
the objects which dominate the structure hierarchy at a given epoch
the ratio of the vorticity to the Hubble parameter, $H$, is roughly
constant and of order a few hundred.  Once an object collapses,
$M\simeq v^2 R/G$ where $M$, $v$, and $R$ are the characteristic mass,
velocity and size of the object.  The spherical collapse model, which
serves as a useful toy-model for structure formation in an expanding
universe, predicts that the mean density within an object is a factor
$f_c\simeq 200$ times the critical density, $\rho_c$.  Moreover,
objects tend to follow the mass-size relation $M\propto R^2$.  If we
put all this together and use the fact that the critical density
scales as the square of the Hubble parameter, we find that
$\omega\simeq 300 H\propto \left (1+z\right )^{3/2}$.  Thus, we expect
larger seed fields at early times.  Similarly, the dynamical time for
objects formed in the early universe is shorter than for objects
today and hence the amplification of the magnetic fields by dynamo
action will be more rapid.  In short, the Biermann battery-dynamo
amplification process may well operate at all stages of the
structure-formation hierarchy.

Various alternatives have also been considered.  For example,
\citet{lazarian:1992} considered a battery driven by electron
diffusion.  \citet{subramanian:1994} considered the Biermann effect
during the epoch of reionization when ionization fronts sweep through
the medium generating currents and magnetic fields.  The fields are
again fairly modest, but could be large enough to seed a dynamo which
then amplifies them to an astrophysically interesting strength.

\subsection{Thermal fluctuations}\label{s-therm}

At finite temperatures, a plasma exhibits a finite (though possibly
low) level of thermal fluctuations at all scales $L = 2\pi/k$ and
frequencies $\omega$.  Differences in the thermal motions between the
different charges generate micro-currents and hence electromagnetic
fields which, on average, provide a noisy electromagnetic background.
In the early universe, the gas is both hot and dilute.  For example,
the gas temperature is $T \sim$ eV both at the recombination epoch and
after reionization but before the formation of LSS.  Therefore, the
cosmic gas constitutes a classical plasma to which the classical
fluctuation theory \citep{sitenko:1967} can be applied.

According to this theory, the power spectral density of magnetic field
fluctuations ${\bf b}$ at wavenumber ${\bf k}$ and frequency $\omega$
is obtained from the spatial correlation function of the fluctuating
magnetic fields and is given by
\begin{equation}
\frac{\langle b_ib_j\rangle_{{\bf k}\omega}}{\sqrt{2\pi}}
= \frac{\mu_0T}{\omega}\left(\delta_{ij}-\frac{k_ik_j}{k^2}\right)
\frac{n^2{\rm Im}\,\epsilon_\perp}{|n^2-\epsilon_\perp|^2}~.
\label{treumann-e01}
\end{equation}
Here, the fluctuating field is designated by lower case letters,
$i,j=1,2,3$, $n^2=k^2c^2/\omega^2$ is the index of refraction and
$\epsilon_\perp(\omega,{\bf k})$ is the complex transverse dielectric
response function of the isotropic plasma. The latter, in
(nonrelativistic) thermal equilibrium, is given by
\begin{equation}
\epsilon_\perp(\omega,{\bf k})=1-\sum_{e,i}\frac{\omega_{e,i}^2}
{2\omega k^2}\int\frac{{\bf v\times(k\times v})F_{0,e,i}'(v)}
{{\bf k\cdot v}-\omega} dv^3,
\label{treumann-e02}
\end{equation}
where the sum is over negative ($e$) and positive ($i$) charges,
$\omega_{e,i}$, the respective plasma frequencies, and
$F_0'\equiv \partial F_0/\partial{\bf v}$, the derivative of the
equilibrium distribution $F_0$ which is assumed to be Maxwellian.
The response function becomes
\begin{equation}
\epsilon_\perp=1-\frac{\omega_e^2}{\omega^2}\left[\Phi(\varpi)+
\frac{\hat \theta}{M}\Phi(M \varpi)-i \varpi \sqrt{\pi}
\left({\rm e}^{-\varpi^2}+\frac{\hat \theta}{M}
{\rm e}^{-M^2 \varpi^2}\right)\right].
\label{treumann-e03}
\end{equation}
Here, $M^2\equiv (m_i/m_e){\hat \theta}$, and $\Phi(\varpi)\approx 2
\varpi^2+\dots$ is the Gordeyev integral of
$\varpi=\sqrt{3/2}(\omega/kc_s)$ with $c_s^2=3T_e/m_e$, the square of
the sound speed, and ${\hat \theta}=T_e/T_i$. Only at very large ion
temperature $T_i\gg T_e$ do the ions contribute to the imaginary part.

Seed fields for a turbulence dynamo arise from the
zero frequency contribution, $\varpi \to 0$ 
{(index 0 in the following equation)}, which yields the
magnetic power spectral density
\begin{equation}
\frac{\langle b_ib_j\rangle_{0{k}}}{\sqrt{2\pi}}\approx
\frac{\mu_0\,m_ec^2}{4\,\omega_e}\frac{{\hat \Theta}^{1/2}\,k\lambda_e}
{(k^2\lambda_e^2-m_e/m_i)^2} \simeq \frac{10^{-37}}
{L_{\rm kpc}{N_{[\mathrm{cm}^{-3}]}}}\sqrt{T_{\rm eV}} 
\quad {\rm \frac{V^2s^3}{m}},
\label{treumann-e04}
\end{equation}
where ${\hat \Theta}\equiv T_e/m_ec^2$ is the normalized temperature 
{$T_e$ and
$T_{e[\mathrm{eV}]}$ is temperature in eV},
$\lambda_e=c/\omega_e\simeq 6/N^{1/2}$ is the electron inertial
length, $N$ is the number density, and {$N_{[\mathrm{cm}^{-3}]}$ 
the density in cm$^{-3}$}.  
To obtain the last
equality we have ignored the term $k^2\lambda_e^2$ which is
negligible on galactic and extragalactic scales.
The magnetic spectral energy density thus grows linearly with
wavenumber $k$, i.e. it decreases with the scale $L$.

For comparison, a $B=1\ \mu$G field has the spectral energy density
$\langle B^2_{1\mu{\rm G}}\rangle_{0k}$ $\approx 4.3\times 10^{-14}$
V$^2$s$^3$/m.  Were we to have integrated over all frequencies {$\varpi$}, 
we would have obtained the equipartition between magnetic and electron
thermal energy densities $\langle b^2\rangle_{\bf k}=2\mu_0\,T_e$
which holds in thermal equilibrium.  Evidently, the thermal
fluctuation level in Equation (\ref{treumann-e04}) is quite low. In
order to obtain a $B= 1\mu$G field, the spectral energy density must
be amplified by the large factor of $\sim 5 \times 10^{23}L_{{\rm
kpc}}{N_{[\mathrm{cm}^{-3}]}}/\sqrt{T_{\rm eV}}$.

\subsection{Filamentation instability}\label{s-fila}

Dynamos on galactic or cluster scales need seed fields on comparably
large scales to get them started.  Moreover, large-scale dynamos
require turbulence and/or helicity which may not exist at adequate
levels.  Recently, the so-called filamentation (or Weibel) instability
\citep{weibel:1959,fried:1959} has been discussed as a generator of
the IGMF \citep{gruzinov:1999,gruzinov:2001,medvedev:1999} and a
possible alternative to dynamos on galactic and extragalactic
scales.  The Weibel instability does require a strong departure of the
plasma state from thermal equilibrium, which can be provided by fast
beams on the plasma background
\citep{achterberg:2007,fried:1959,sakai:2004} or temperature
anisotropies.  The latter can be caused by pressure anisotropies
\citep{weibel:1959,treumann:2010}, shock waves
\citep{jaroschek:2005,nishikawa:2009}, or thin current sheets.
Turbulence, if it exists in the early universe, leads to thin current
sheets which can accelerate particles by second-order Fermi
acceleration \citep{jaroschek:2008,jaroschek:2009}, the local electric
field which is generated in the current sheets, or by reconnection.
Here, we discuss the possibility that the instability is driven by
kinetic pressure/temperature anisotropies in the absence of magnetic
field, as a mechanism to create seed fields.

Pressure and temperature anisotropies are probably superior to the
beam for driving the instability for the simple reason that beams are
highly unstable with respect to high-frequency plasma instabilities
(Langmuir and Buneman modes).  If the beams have velocity $V_b$ and
are current-compensated (i.e., the velocities of particles with
different charges are the same) then the two-stream instability is
excited with growth rate $\gamma_{ts}\simeq
\sqrt{3}(N_b/2N)^{1/3}\omega_{e}$.  Current-compensated beams arise in
the symmetric counter-streaming configurations found in shocks.  For
the reasonable beam density $N_b\sim 0.1 N$ we find
$\gamma_{ts}\approx 0.6\,\omega_{e}\sim $ few kHz with a wavenumber is
$k_{ts}\sim \omega_e/V_b$.  If the beams are not current compensated,
which is the case in turbulence where a multitude of thin current
sheets is generated, then the Buneman instability is excited with
growth rate $\gamma_B\sim 0.03\,\omega_e\sim 10^2$ Hz, if the current
velocity $V_c=V_e-V_i>v_e$ exceeds the electron thermal speed $v_e$.
Otherwise, for $V_c<v_e$, the ion sound will be excited with growth
rate roughly $\gamma_{ia}\sim\omega_pi\sim$ few 10$^2$ Hz.  All these
instabilities grow very fast and readily deplete the beams/streams,
heating the electron plasma in the direction of flow and thus causing
the pressure or temperature anisotropy of electrons with higher
temperature along the flow direction, i.e. $||$ to ${\bf V}_c$. If
this is assumed to be the (readily achieved) final microscopic state,
then the temperature anisotropy $A=T_{e\|}/T_{e}-1>0$ becomes unstable
with respect to the Weibel thermally anisotropic mode. The temperature
anisotropy corresponds to a pressure anisotropy ${\sf P}_e=N[T_e{\sf
I}+(T_{e\|}-T_e){\bf V}_c{\bf V}_c/V_c^2]$.  One should note that
only the electrons, because of their large mobility, are important;
the ions are much less active on these microscopic time scales and
thus do not contribute.

In this thermally anisotropic case, the electrons obey a
bi-Maxwellian equilibrium distribution function
\begin{equation}
f_e(v_\perp,v_\|)=\frac{(m_e/2\pi )^{3/2}}{T_e\sqrt{T_{e\|}}}
\exp\left[-\frac{m_ev_\perp^2}{2T_e}-\frac{m_ev_\|^2}{2T_{e\|}}\right],
\label{treumann-e07}
\end{equation}
and the Weibel instability takes over with the linear
electromagnetic dispersion relation $n^2=\epsilon_{\perp}$
and the transverse dielectric response function
\begin{equation}
\epsilon_\perp\equiv 1-\frac{\omega_e^2}{\omega^2} \left\{ 1-(A+1)
\left[ 1+\zeta Z(\zeta) \right] \right\} -\frac{\omega_i^2}{\omega^2}
= n^2.
\label{treumann-e08}
\end{equation}
Here, $Z(\zeta)$ is the plasma dispersion function,
$\zeta=\omega/k_\perp v_{e}$, and $v_{e}=\sqrt{2T_e/m_e}$ is
the electron thermal speed perpendicular to current.
The Weibel instability grows in the plane perpendicular
to the direction of higher thermal velocity, which in our case
has been assumed as the parallel direction. Hence,
${\bf k}\equiv{\bf k}_\perp$. The contribution of the resting ions
has been retained for completeness; because of the smallness of
the ion plasma frequency $\omega_{\,i}\ll\omega_{\,e}$, being
much less than the electron plasma frequency $\omega_{\,e}$,
it plays no role in the instability but is important in
the discussion of the thermal level.

\begin{figure}
\vskip 0cm\hskip 1.5cm
\includegraphics[width=0.7\textwidth,height=0.7\textwidth]{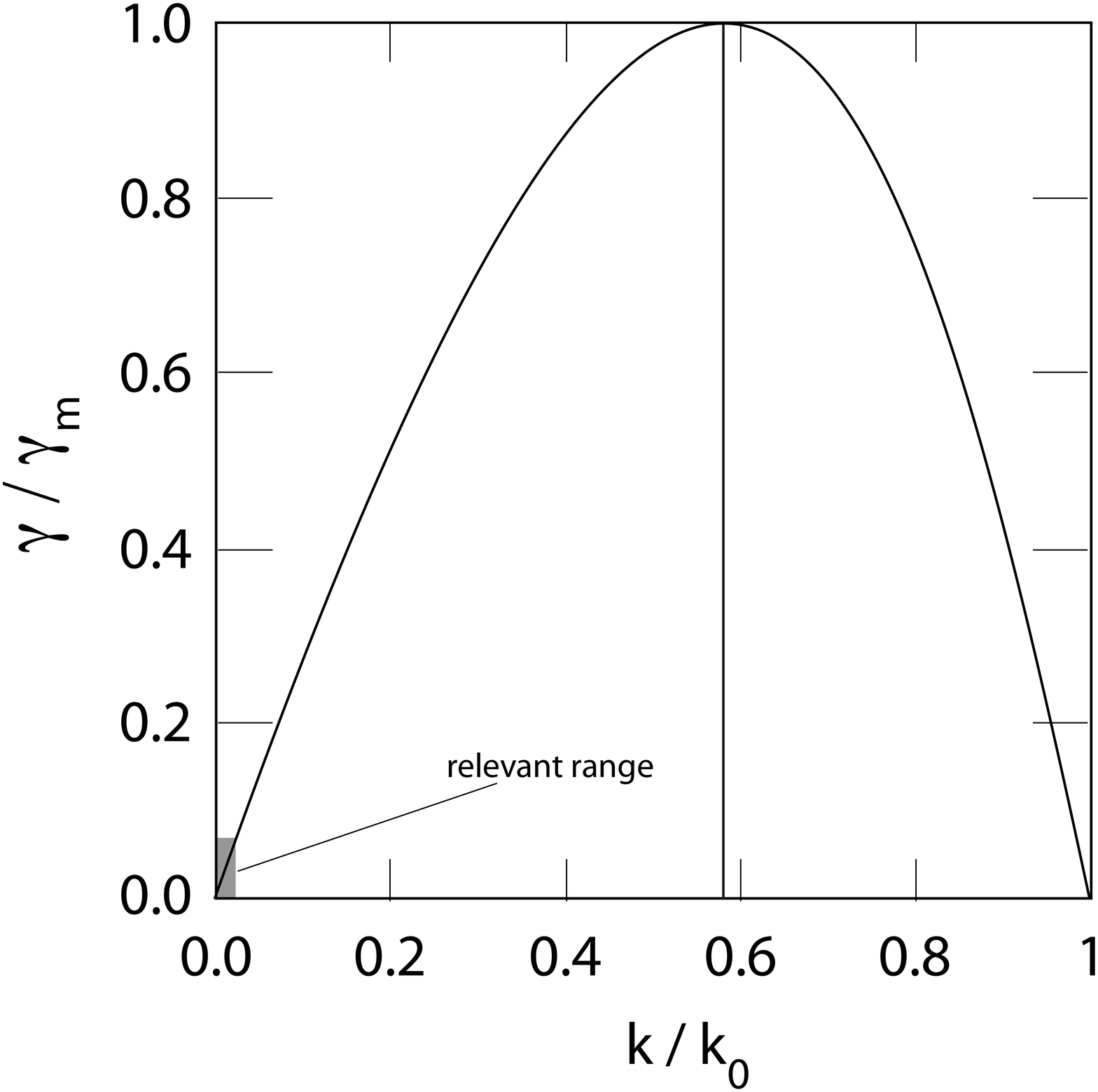}
\vskip -0.1cm
\caption{The growth rate {$\gamma=\gamma_W$} of the anisotropic-thermal 
Weibel instability normalized to the maximum growth, $\gamma/\gamma_m$,
as a function of the normalized wavenumber $k/k_0$ (see the text for 
the definition
of $k_0$). The vertical
line indicates the position of the maximum growing wavenumber
{$k_m/k_0 =1/\sqrt{3}$}.}
\label{treumann-f1}
\end{figure}

At zero real frequency, $\omega=i\gamma$, and the instability
$\gamma(k_\perp)>0$ sets on for phase velocities
$\omega/k_\perp\ll v_{e}$ and wavenumbers $k_\perp<k_{\rm 0}$ with
growth rate
\begin{equation}
\frac{\gamma_{\,W}}{\omega_e}\simeq \sqrt{\frac{2}{\pi}}
\frac{v_{e\perp}}{c}\frac{k_\perp}{k_0}\left(1 -
\frac{k_\perp^2}{k_{\rm 0}^2}\right)(A+1)(k_{\rm 0}\lambda_e)^3,
\label{treumann-e09}
\end{equation}
vanishing at infinite wavelength $k_\perp=0$, where
$k_{0}\lambda_e= \sqrt{A}$. The growth rate maximizes at
wavenumber $k_{{m}}\lambda_e=k_0\lambda_e/\sqrt{3}=\sqrt{A/3}$
(see Figure \ref{treumann-f1}), where its value is
\begin{equation}
\gamma_{\,m}\approx 34 \sqrt{N_{\rm [cm^{-3}]}T_{e[{\rm eV}]}}
A^{3/2}(A+1) \ \ {\rm Hz}.
\label{treumann-e10}
\end{equation}
A substantial growth can be achieved and stationary (purely growing)
non-fluctuating, short-wavelength magnetic fields are produced. The
maximally growing wavelength is indeed very short. For a plasma
density of $N\approx 0.01$ cm$^{-3}$, the wavelength is only of the
order of $\lambda=(2\pi c/\omega_e)\sqrt{3/A}\sim 50\sqrt{3/A}$ km.
Even very weak anisotropies of $A\sim 10^{-4}$ give $\lambda_m\sim
10^4$ km. Galactic or intergalactic scales {$L$} are far longer. For a
realistic anisotropy of $A\ \la\ 1$, at these large scales the growth
rate scales with $L$ as
\begin{equation}
\gamma_W\simeq4A\sqrt{\pi{\hat \Theta}}(c/L)\simeq 10^{-13}
A\sqrt{T_{\rm eV}}/L_{\rm kpc}\quad {\rm Hz}
\label{treumann-e11}
\end{equation}

The above tells that mostly magnetic fields of small scales are
excited and saturated by the instability, and those of galactic or
intergalactic scales would not be substantial. Yet for $A \sim 0.1$
and $T \sim$ eV, the growth rate is $\gamma_W\sim 10^{-14}$ Hz on
kpc-scales which is quite large.  Magnetic fields of $\sim 10^{-16}$ G
on kpc-scales would arise in $\sim 10^7$ years or so provided the
fields on smaller scales do not saturate the instability.

\section{Magnetic Fields after Recombination}\label{s-recom}

If magnetic fields were created before or during recombination they
could have had a significant impact on the thermal and chemical
evolution of gas during the dark ages, on the formation of first stars
and galaxies, and on the epoch of reionization
\citep[e.g.,][]{ss05,Tashiro06a,Sethi08,Schleicher08b,Schleicher09prim}
\citep[see also][of this volume]{wrss10}.   On the other hand, the
first stars and galaxies may have been a source of magnetic fields.
Our discussion should be viewed with a critical eye since the
existence, strength and scale of early universe fields is highly uncertain
\citep[see e.g.][]{Grasso01} as is the efficiency of dynamo action at
early times.

\subsection{Magnetic fields during the dark ages}\label{s-dark}

The dark ages designate the period between recombination and
reionization, during which no astronomical objects exist.  The only
radiation produced during this epoch is the 21 cm line of neutral
hydrogen.  The primordial gas consists of about $75\%$ hydrogen and
$25\%$ helium, as well as traceable amounts of lithium and beryllium
(both of the order $10^{-9}$) \citep[see, e.g.,][]{Kolb90}. The
ionization degree freezes out at about $10^{-4}$ \citep{Peebles68,
Zeldovich69, Seager99}. Though this ionization degree may seem
rather low, it is significantly higher than the typical ionization
degree in present-day molecular clouds \citep{Myers95}.  More to the
point, it is high enough to prevent non-ideal magnetohydrodynamic
(MHD) processes such as ambipolar or Ohmic diffusion from dissipating
the magnetic fields.

The more relevant question is whether magnetic fields can be sustained
during gravitational collapse when both the density and recombination
rate increase and hence the ionization degree
decreases.  Gravitational collapse occurs on length scales where the
thermal pressure can no longer balance the gravitational force. The
critical length scale is given by the Jeans length
\begin{equation}
\lambda_J=\left ( \frac{\pi c_s^2}{G\rho} \right)^{1/2},
\label{schleicher-e01}
\end{equation}
where $c_s$ is the thermal sound speed and $\rho$ the mass
density. Previous studies by \citet{Maki04} and \citet{Glover09}
followed the chemical evolution during gravitational collapse.  Their
one-zone model takes into account the density evolution in
the central core, where the collapse takes place on the free-fall timescale
$t_{ff}\sim1/\sqrt{G\rho}$. As is well-known from analytic solutions and
numerical simulations, the density in the central core is roughly
homogeneous on the scale of the Jeans length $\lambda_J$
\citep{Larson69, Penston69, Abel02, Bromm03, Yoshida08}.

The calculations of \citet{Maki04} and \citet{Glover09} showed
that the ionization degree evolves roughly as $\rho^{-1/2}$
until a density of $\sim10^9$ cm$^{-3}$ is reached. At this stage,
the recombination rate becomes so high that the proton abundance
becomes negligible. However, the presence of lithium and
the low recombination rate of Li$^+$ stabilizes the ionization
degree at a level of $\sim10^{-10}$, and it stays roughly
constant even at higher densities.

In order to assess the implications of this result,
\citet{Schleicher09prim} combine the chemical network of
\citet{Glover09} with a state-of-the-art model for ambipolar and Ohmic
diffusions, based on the work of \citet{Pinto08a} and
\citet{Pinto08b}. In their multi-fluid approach, the ambipolar
diffusion rate is given as
\begin{equation}
L_{\rm{AD}}=\frac{\eta_{\rm{AD}}}{4\pi}\left|\left(\nabla\times
\vec{B}\right)\times\vec{B}/B\right|^2,
\label{schleicher-e02}
\end{equation}
where the $\vec{B}$ denotes the magnetic field. The ambipolar
diffusivity $\eta_{\rm{AD}}$ is given as
\begin{equation}
\eta_{\rm{AD}}^{-1}=\sum_i \eta_{{\rm AD},i}^{-1},
\label{schleicher-e03}
\end{equation}
with $\eta_{{\rm AD}, i}$ denoting the ambipolar diffusivity
due to an ionized species $i$. Similarly the Ohmic diffusion
rate is given as
\begin{equation}
L_{\rm{Ohm}}=\frac{\eta_O}{4\pi}\left | \nabla\times\vec{B}
\right|^2,
\label{schleicher-e04}
\end{equation}
with the Ohmic resistivity $\eta_O$ given as
\begin{equation}
\eta_{\rm{O}}^{-1}=\sum_i \eta_{{\rm O},i}^{-1}.
\label{schleicher-e05}
\end{equation}
The ambipolar resistivities $\eta_{{\rm AD}, i}$ and the Ohmic
resistivities $\eta_{{\rm O},i}$ due to ionized species $i$
are calculated from the momentum-transfer coefficients as
described in the appendix of \citet{Pinto08a}.

To guide the physical intuition, we briefly summarize their results for 
the case of two charged fluids, e.g. electrons and positive ions, but note 
that the more detailed multi-fluid approach has been adopted in the 
calculations. The key quantities that regulate non-ideal MHD effects 
are the Hall parameters $\beta_{\mathrm{sm}}$, defined as
\begin{equation}
\beta_{\mathrm{sn}} = \left( \frac{q_s B}{m_s c} \right)\frac{m_s+m_n}
{\rho_n \langle \sigma v\rangle_{\mathrm{sn}}}.
\end{equation}
Here, $q_s$ denotes the charge of the species $s$, $m_s$ its mass, 
$m_n$ the mass of the neutrals, $\rho_n$ the neutral mass density, $c$ 
the speed of light and $\langle \sigma v\rangle_{sn}$ the momentum-transfer 
coefficients between the ions and the neutrals (for the expressions, see 
\citet{Pinto08b} and the discussion below). As shown by \citet{Pinto08a}, 
the ambipolar and Ohmic resistivities are then given as
\begin{eqnarray}
\eta_{\mathrm{AD}}&=&\left(\frac{c^2}{4\pi \sigma} \right)
\frac{\beta_{+n}|\beta_{-n}|}{\beta_{+n}+|\beta_{-n}}|,\\
\eta_{\mathrm{O}}&=&\left( \frac{c^2}{4\pi \sigma}  \right)
\frac{1}{\beta_{+n}+|\beta_{-n}|},
\end{eqnarray}
where $\beta_{+n}$, $\beta_{-n}$ denote the Hall parameters for 
the positively and negatively charged species, and $\sigma$ is given as
\begin{equation}
\sigma=\frac{q_+ \rho_+ c}{m_+ B}.
\end{equation}
In the latter expression, $q_+$ denotes the positive charge, and $\rho_+$ 
the mass density of the positively charged species.

For collisions
involving protons and electrons, the momentum-transfer
coefficients of \citet{Pinto08b} are adopted. For collisions
involving Li$^+$, the momentum-transfer coefficients using
the polarization approximation are calculated, as described
by \citet{Schleicher09prim}. This is indeed justified for
collisions with helium \citep{Cassidy85} and H$_2$
\citep{Dickinson82,Roeggen02}. For collisions with atomic
hydrogen, no detailed theoretical or experimental measurements
are currently available. This process is of minor importance,
however, as Li$^+$ becomes the dominant ionized species at
densities  when molecular hydrogen is the dominant neutral species.

\begin{figure}
\vskip -0.3cm\hskip 1.3cm
\includegraphics[width=0.7\textwidth,height=0.8\textwidth]{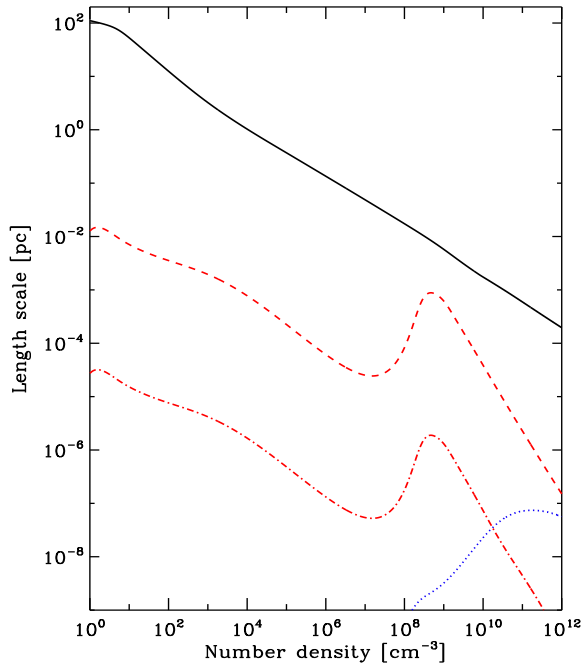}
\vskip -0.1cm
\caption{Diffusion length scales during gravitational collapse.
The black solid line denotes the Jeans length as a function of
central core density. The ambipolar diffusion scale for the
equipartition magnetic field is given by the red dashed line,
while the red dashed-dotted line corresponds to the scale for a
weaker field (the magnetic energy equals $1/60$ of the
kinetic energy). The blue dotted line gives the Ohmic diffusion
scale, which is independent of the magnetic field strength.
In all cases, the dissipation scales are smaller than
the Jeans length.}
\label{schleicher-f1}
\end{figure}

Equations (\ref{schleicher-e02}) and (\ref{schleicher-e04})
show that the dissipation rates depend on the magnetic field
strength itself, and we note that also the ambipolar resistivities
increase with increasing field strength \citep{Pinto08a}.
During gravitational collapse, an initially weak field, for
instance due to the Biermann battery \citep{xocn08}, may be amplified
both due to gravitational compression as well as turbulence dynamo
\citep{Schleicher10c}. If the amplification due to turbulence dynamo
is efficient, saturation may occur at equipartition or a somewhat
lower level \citep[see, e.g.,][]{Subramanian99,cv00,Haugen04a,Haugen04b,
Haugen04c,Schekochihin04,Brandenburg05,cvbl09}.
As ambipolar and Ohmic diffusions are stronger for stronger
fields, we will initially focus on equipartition magnetic
fields and show that magnetic fields can be sustained. The same is
then true also for weaker magnetic fields.

To have a definite model, we assume Kolmogorov turbulence, and
thus that the equipartition field strength scales as $l^{1/3}$.
This is a good approximation for subsonic turbulence where
compressional effects are of minor importance, and it is this
type of turbulence that is expected in the first star-forming
systems \citep{Abel02}. The Ohmic and ambipolar diffusion scales
are calculated as a function of the central core density and
compared them to the Jeans length (see Figure \ref{schleicher-f1}).
For clarity, the Ohmic / ambipolar diffusion scales are defined as 
the length scales on which the Ohmic or ambipolar diffusion time equals 
the eddy-turnover time. The calculation assumes that the turbulent 
velocity and the magnetic field follow a typical scaling law as expected 
for Kolmogorov turbulence. During collapse, the increased densities 
lead to a more efficient coupling between the ions and the neutrals, 
which largely compensates for the decreasing ionization degree.
At a density of $10^9$ cm$^{-3}$, the ambipolar diffusion scale increases
due to a drop of the ionization degree at these high densities,
but still stays an order of magnitude below the Jeans length. The
Ohmic diffusion scale always stays orders of magnitude below the
Jeans length. The figure also shows that when the magnetic field
strength is below the equipartition value, the ambipolar diffusion
scale is even small. The Ohmic dissipation scale, however, stays
on the same low level, as the Ohmic dissipation rate scales
roughly with the magnetic energy.

Numerical calculations suggest that the ionization degree may
even be higher than expected from one-zone models, in particular
due to the presence of shocks of Mach number $\sim 1$
\citep{Clark10, Turk10}. So we safely conclude that during the
dark ages, the magnetic fields up to the equipartition strength
can be sustained even at high densities in collapsed regions.

\subsection{Magnetic field evolution in first star-forming halos}
\label{s-halos}

We now turn to the amplification of magnetic fields during the
formation of the first stars.  The halos that harbor the first stars
at redshifts of $z \sim 20-30$ typically have a total mass of $10^6\
M_\odot$, consist of primordial gas with a temperature of $\sim3000$
K, and have a spatial extent of $\sim100$ pc. Turbulence is expected
to be present, but on a subsonic level \citep{Abel02, Bromm03,
Yoshida08}. Such turbulence may create strong magnetic
fields in the first galaxies and around them \citep{rkcd08,
Arshakian09, deSouza10, Schleicher10c}.

Consider a simple model in which magnetic fields are
amplified and saturate at a level corresponding to an
energy $1/60$ of the equipartition energy on an eddy turnover
time $t_{\rm{eddy}}$. We assume that turbulence is injected during
the virialization phase on the length scale of the system
(i.e., $\sim100$ pc) with the velocity of the order of the sound speed.
Both Kolmogorov and Burgers turbulences are considered; for
Kolmogorov turbulence the velocity depends on the length scale
as $l^{1/3}$, while for Burgers turbulence the scaling accords
to $l^{1/2}$.

\begin{figure}
\vskip -0.4cm\hskip 0.2cm
\includegraphics[scale=0.6]{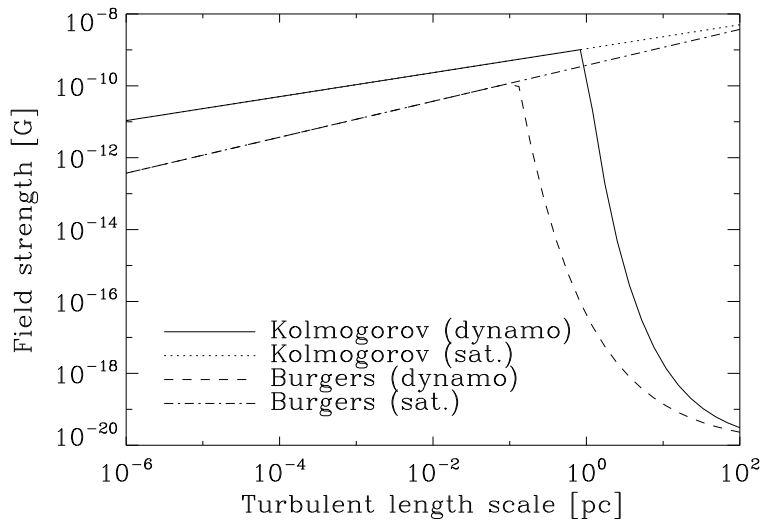}
\vskip -0.2cm
\caption{The magnetic field strength after a free-fall time. 
The horizontal axis denotes the length scale of the turbulent eddies. 
We assume a power-law scaling relation for the rms turbulence on a given 
length scale, and explore the cases of Kolmogorov and Burgers-type 
turbulence (Schleicher et al. 2010). In this model, the turbulent 
injection scale is $100$~pc. For every length scale, we show both 
the maximum field strength that could be obtained from the small-scale 
dynamo, as well as our estimate for the actual field strength that might 
be produced within a free-fall time. The model suggests the production 
of strong magnetic fields on small scales.}
\label{schleicher-f2}
\end{figure}

For magnetic fields to be relevant during the formation of the first
stars, they must be amplified within the free-fall time scale
$t_{ff}\sim1/\sqrt{G\rho}$.  The number of eddy turnovers is thus
given by the ratio $t_{ff}/t_{\rm{eddy}}$. Here, the eddy turnover
time $t_{\rm{eddy}} \equiv l/v$ has a characteristic dependence on
length scale for Kolmogorov and Burgers turbulences.  Then, the
expected magnetic field strength is given in Figure
\ref{schleicher-f2} as a function of the length scale of
turbulence. The preliminary calculation shows that magnetic fields of
the order $10^{-9}$ G can be reached in the first star-forming
halos. For a comprehensive understanding of the strength and structure
of the magnetic fields, the detailed implementation of turbulence
dynamo will be required.
Recent numerical MHD simulations suggest that the small-scale dynamo 
indeed operates during gravitational collapse, producing magnetic 
fields during the formation of the first structures
\citep{ssbf10,fssb11}.

Additional amplification may occur after the formation of protostellar
disks.  In particular, a large-scale dynamo and the magneto-rotational
instability may further enhance the magnetic fields in the first
star-forming halos \citep{Pudritz89,Tan04,Silk06}.  For simplified
field geometries, \citet{Machida06,Machida08} found that magnetic
fields lead to the formation of jets and help to suppress
fragmentation in protostellar disks.  \citet{Fromang04} studied
self-gravitating magnetized disks and found that the interaction of
turbulence created by the magneto-rotational instability may excite
additional modes for the gravitational instability, and that the
interaction of these modes reduces the accretion rate.  This work
suggests that magnetic fields may be amplified rapidly in the first
star-forming halos and may become dynamically relevant though more
realistic studies are necessary.

\subsection{Seed fields from astrophysical processes}\label{s-sfao}

The first stars probably possessed strong magnetic fields and
therefore may have provided seed fields for dynamos in galaxies and in
the IGM.  If the stars subsequently explode as supernovae or lose a
significant amount of mass through stellar winds, the fields ejected
along with mass will find their way into the interstellar medium (ISM)
and spread beyond galaxies into the IGM through galactic winds.
Simple estimates by \citet{syro70} illustrate the viability of the
process.  If there have been some $10^8$ supernovae over the lifetime
of galaxies, each of which spreads material through a $\left(10\ \rm
pc\right)^3$ volume.  Using values for the field strength typical of
the Crab nebula, one therefore expects galaxies to be filled by 10 pc
regions with fields of strength $\sim 3\ \mu$G.  Assuming the same
$L^{-3/2}$ scaling, one finds fields of strength of $\sim 10^{-11}$ G
on 10 kpc scales.  This value is significantly larger than those from
the processes described in Section \ref{s-plas} and in \citet{wrss10}
of this volume, although the filled volume is rather small.  Recently,
\citet{ddlm09} suggested that galactic outflows during the starburst
phase of galactic evolution can deposit a substantial amount of
magnetic fields in the IGM.

Seed fields can be produced at cosmological shocks which were induced
during the formation of the LSS of the universe
\citep{rkhj03,psej06,krco07,sohb08,vbg09} (see Section \ref{s-turb} for
further description of cosmological shocks).
Cosmological shocks are collisionless like shocks in other astrophysical
environments, where CRs are accelerated at the same time as the gas is
heated.
During the process of acceleration, it was shown that the upstream
magnetic field can be amplified nonlinearly by non-resonant
growing modes \citep{bell04}.
Then, the magnetic field can have the energy up to
\begin{equation}
\varepsilon_B \sim \frac{1}{2}\frac{U_{n1}}{c}\varepsilon_{\rm CRs}.
\label{ryu-e12}
\end{equation}
With $U_{n1}/c \sim 10^{-3}$ and Mach number a few in cosmological shocks
\citep{rkhj03,krco07}, we get
$\varepsilon_B \sim \varepsilon_{\rm CRs} \sim \varepsilon_{\rm therm}$.
In addition, the Weibel instability described in Section \ref{s-fila}
can operate and produce magnetic fields up to the level of
$\varepsilon_B \sim 10^{-3} \varepsilon_{\rm sh}$ \citep{ss03,msk06}.
Here, $\varepsilon_{\rm sh}$ is the energy density of upstream flow,
and $\varepsilon_B$, $\varepsilon_{\rm CRs}$, and $\varepsilon_{\rm therm}$
are the energy densities of downstream magnetic fields, CRs, and gas
random motion, respectively.
These processes can potentially produce strong fields around cosmological
shocks, although the volume filling is small and the coherence length of
the resulting fields is expected to be microscopic.

Other mechanisms have been proposed.  For instance, recently,
\citet{mb10} suggested that the return current which is induced by
cosmic-rays produced by early supernovae can deposit seed fields into
the IGM.

\section{Magnetic Fields in the IGM}\label{s-ampl}

With the processes described in Sections \ref{s-plas} and
\ref{s-recom} and also in \citet{wrss10} of this volume, there is no
shortage of mechanisms to generate seed fields for the IGMF.  Those
fields are expected to be amplified by the turbulent flow motions
which were induced during the formation of the LSS of the universe.
The turbulence dynamo not only increases the strength of magnetic
fields, but also produces the magnetic fields of large scales, up to
the energy injection scale, through the inverse cascade.  The beauty
of the turbulence dynamo in LSS is that it erases the memory of weak
seed fields and produces the IGMF, independent of the origin of seed
fields.

\subsection{Turbulence in the LSS of the universe}\label{s-turb}

Signatures of turbulence have been observed in the ICM.  For example,
\citet{sfmb04} analyzed gas pressure maps that were constructed from
XMM-Newton X-ray data.  They claimed that in the Coma cluster, which
appears to be in a post-merger state, pressure fluctuations are
consistent with Kolmogorov turbulence.  The turbulence is likely
subsonic but with an energy that is at least 10 \% of the thermal
energy, i.e., $\varepsilon_{\rm turb} > 0.1 \varepsilon_{\rm therm}$.
The results agree with predictions from numerical simulations, namely
that the flows in cluster scales have a power spectrum expected for
Kolmogorov turbulence \citep{kcor97,kz07}, and even in relaxed
clusters the flow motions have $\varepsilon_{\rm kin} \sim 0.1
\varepsilon_{\rm therm}$ \citep{nvk07}.  Turbulence in the ICM was
also studied in RM maps of a few clusters \citep{ve05,gmgp08,bfmg10}.

Recently, \citet{rkcd08} proposed a scenario in which vorticity is
generated directly or indirectly at cosmological shocks and turbulence
in the IGM is induced via the cascade of the vorticity.
Here, we provide an estimate estimate of the
turbulence seen in simulations for the formation of the LSS.  The
results are based on a simulation described in \citet{co06} which
includes the radiative processes of heating/cooling and feedback from
galactic superwind.  The work of \citet{rkcd08} utilized a simulation,
where only the gravitational and gas dynamical processes are included.

In the simulation of \citet{co06}, the WMAP1-normalized $\Lambda$CDM
cosmology was employed with the following parameters:
$\Omega_b=0.048$, $\Omega_m=0.31$, $\Omega_{\Lambda}=0.69$, $h \equiv
H_0$/(100 km/s/Mpc) = 0.69, $\sigma_8 = 0.89$, and $n=0.97$.  A cubic
box of comoving size $85\ h^{-1}$ Mpc was simulated using $1024^3$
grid zones for gas and gravity and $512^3$ particles for dark matter.
It allows a uniform spatial resolution of $\Delta l = 83\ h^{-1}$ kpc.
For detailed descriptions of input physics ingredients such as
non-equilibrium ionization/cooling, photoionization/heating, star
formation, and feedback processes, refer \citet{copw03} and
\citet{co06}.  Feedback from star formation was treated in three
forms: ionizing UV photons, galactic superwinds, and metal enrichment.
Galactic superwinds were meant to represent cumulative supernova
explosions, and modeled as outflows of several hundred km s$^{-1}$.
The input of galactic superwind energy for a given amount of star
formation was determined by matching the outflow velocities computed
for star-burst galaxies in the simulation with those observed in the
real world.  The simulations were performed using a PM/Eulerian
hydrodynamic cosmology code \citep{rokc93}.

\begin{figure}
\vskip 0cm\hskip 0cm
\includegraphics[width=1\textwidth]{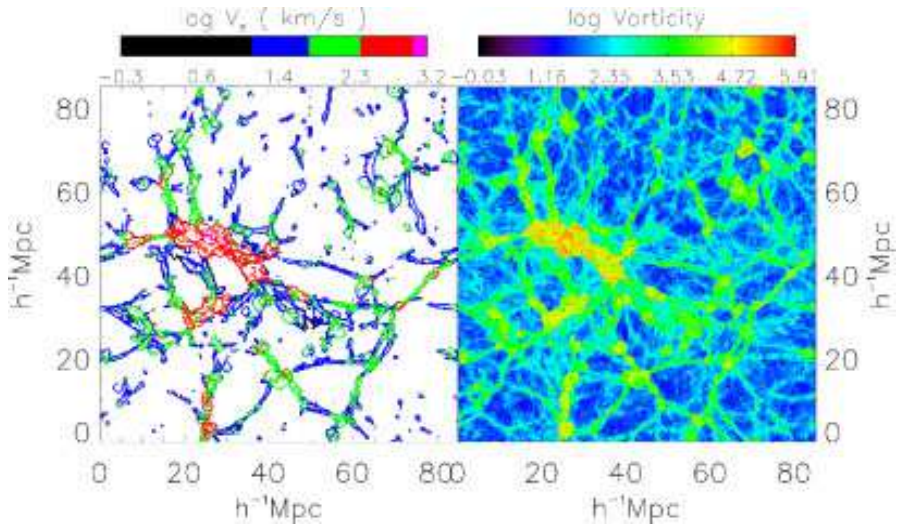}
\vskip 0cm
\caption{Two-dimensional slice of $(85\ h^{-1}{\rm Mpc})^2$ showing
shock locations with color-coded shock speed (left panel) and the
magnitude of vorticity (right panel) at $z=0$.
The vorticity is given in units of $10^{-4}\ t_{\rm age}^{-1}$,
where $t_{\rm age}$ is the age of the universe.}
\label{ryu-f1}
\end{figure}

In the IGM, vorticity, ${\vec \omega} \equiv {\vec\nabla}\times{\vec v}$,
can be generated directly at curved shocks and also by the baroclinity
of flows.
For uniform upstream flow, the vorticity produced behind curved
shock surface is
\begin{equation}
{\vec \omega}_{\rm cs} = \frac{(\rho_2 - \rho_1)^2}{\rho_1 \rho_2}
K {\vec U}_1 \times {\hat n},
\label{ryu-e01}
\end{equation}
where $\rho_1$ and $\rho_2$ are the upstream and downstream gas
densities, respectively, ${\vec U}_1$ is the upstream flow velocity
in the shock rest frame, $K$ is the curvature tensor of the shock
surface, and ${\hat n}$ is the unit vector normal to the surface.
If isopycnic surfaces do not coincide with isobaric surfaces,
vorticity is generated with the rate given by
\begin{equation}
{\dot{\vec \omega}}_{\rm bc} = \frac{1}{\rho^2}
{\vec\nabla}\rho\times{\vec\nabla}p.
\label{ryu-e02}
\end{equation}

Shock waves are ubiquitous in the IGM, as in other astrophysical
environments.
The spatial distribution and properties of cosmological shocks in
the LSS of the universe have been studied quantitatively using
simulations for the formation of LSS
\citep{rkhj03,psej06,krco07,sohb08,vbg09}.
In the cold dark matter universe with cosmological constant
($\Lambda$CDM), shocks with Mach number up to a few hundreds and
speed up to a couple of thousand km s$^{-1}$ are present at the
present universe $(z=0)$.
In the left panel of Figure \ref{ryu-f1}, the spatial distribution
of cosmological shocks is shown.
Numerous shocks are found.
External shocks exist around sheets, filaments, and knots of mass
distribution, which form when the gas in void regions accretes
onto them.
Within those nonlinear structures, internal shocks exist, which
form by infall of previously shocked gas to filaments and knots,
and during subclump mergers, as well as by chaotic flow motions.
Due to the low temperature of the accreting gas, the Mach number
of external shocks is high, extending up to $M \sim$ a few
$\times 100$, while internal shocks have mostly low Mach number
of $M \sim$ a few.
The mean distance between shock surfaces is
$\sim 3\ h^{-1}{\rm Mpc}$ when averaged over all the universe,
or $\sim 1\ h^{-1}{\rm Mpc}$ inside nonlinear structures.
Internal shocks of $M \sim 2 - 4$ formed with hot and high-density
gas are responsible for most of shock dissipation into heat and CRs.
It was shown that the shock dissipation can count most of the gas
thermal energy in the IGM \citep{krcs05}.

\begin{figure}
\vskip -3.8cm\hskip -1.5cm
\includegraphics[width=1.3\textwidth]{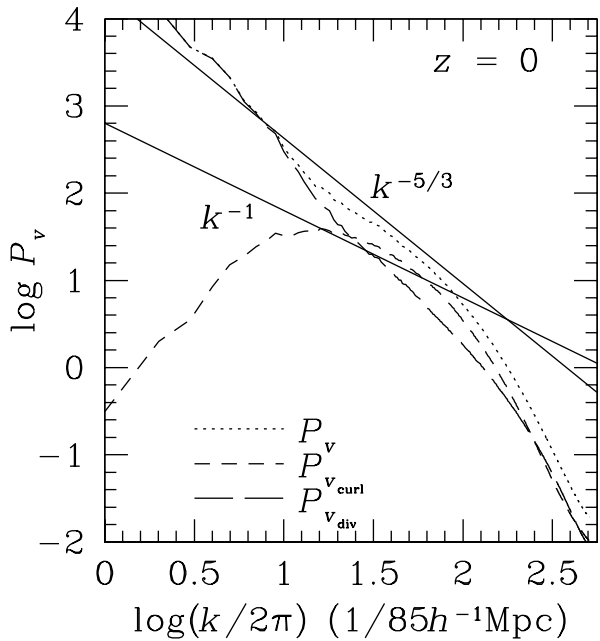}
\vskip -2.8cm
\caption{Power spectra, $\int P_v dk = \left<(1/2) v^2 \right>$, for
the gas velocity and its curl and divergence components at $z=0$.
Two straight lines of slopes $-5/3$ and $-1$ are also drawn for
comparison.}
\label{ryu-f2}
\end{figure}

In the right panel of Figure \ref{ryu-f1}, the distribution of
vorticity is shown.
It closely matches that of shocks, suggesting that a substantial
portion of the vorticity, if not all, has been generated at
the shocks.
As a matter of fact, as was noted in \citet{rkcd08}, the vorticity in
the IGM can be accounted with that generated either directly at curved
cosmological shocks or by the baroclinity of flows.
The contributions from the two processes are comparable.
The baroclinity resulted from the entropy variation induced at shocks.
So all the vorticity generation also can be attributed to the presence
of cosmological shocks.

\begin{figure}
\vskip 0cm\hskip 0cm
\includegraphics[width=1\textwidth]{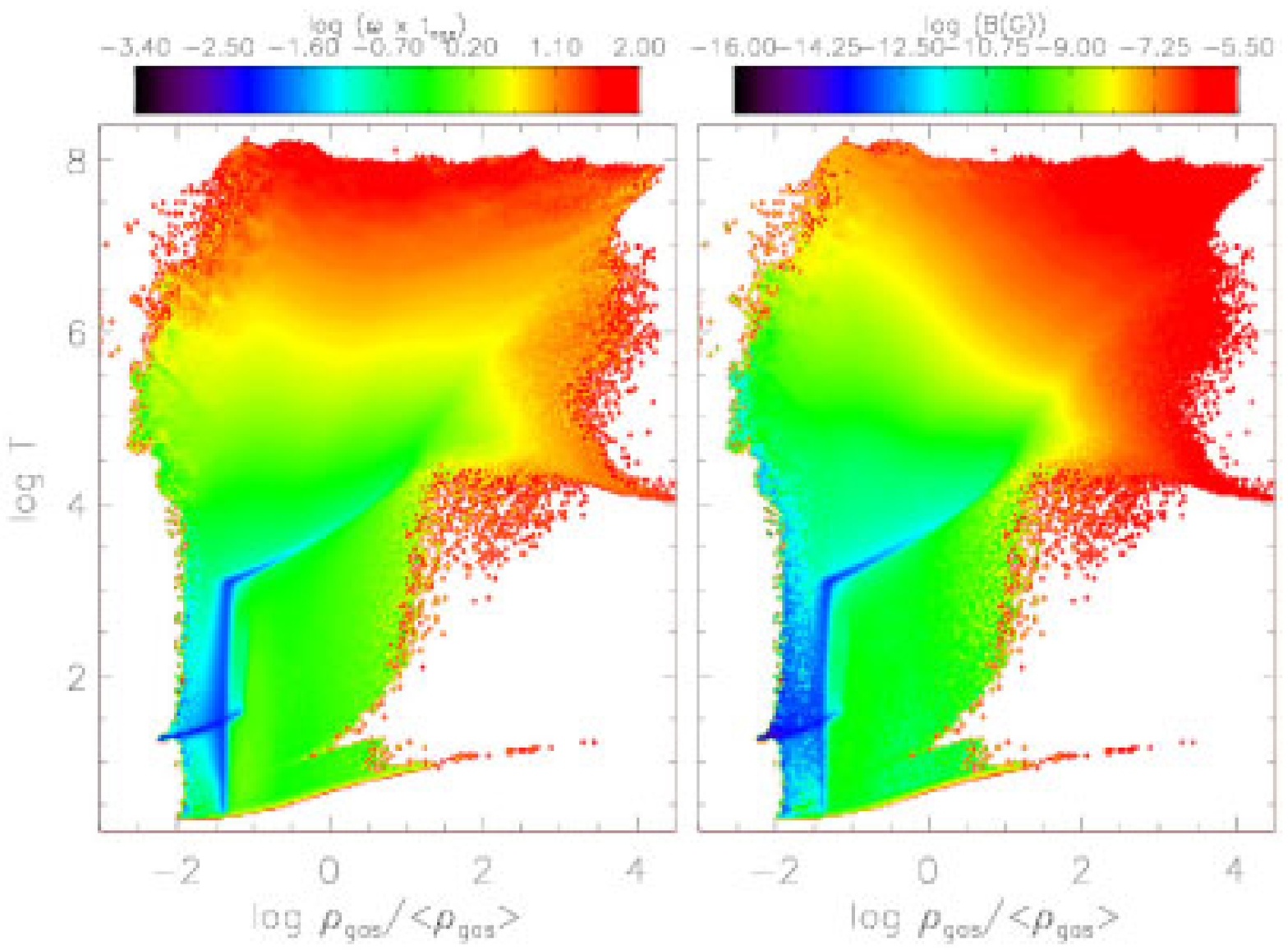}
\vskip 0.1cm
\caption{Distribution of the magnitude of the vorticity in the IGM
(left panel) and the strength of the IGMF (right panel) in the
density-temperature plane at $z=0$.
The vorticity is given in units of $t_{\rm age}^{-1}$.}
\label{ryu-f3}
\end{figure}

For quantification of vorticity, the flow velocity is decomposed into
\begin{equation}
{\vec v} = {\vec v}_{\rm div} + {\vec v}_{\rm curl} +{\vec v}_{\rm unif},
\label{ryu-e03}
\end{equation}
where the divergence and curl components are defined as
${\vec\nabla}\cdot{\vec v}_{\rm div} \equiv {\vec\nabla}\cdot{\vec v}$
and
${\vec\nabla}\times{\vec v}_{\rm curl} \equiv {\vec\nabla}\times{\vec v}$,
respectively.
That is, ${\vec v}_{\rm div}$ is associated to compressional motions,
while ${\vec v}_{\rm curl}$ to incompressible shear motions.
Here ${\vec v}_{\rm unif}$ is the component uniform across the
computational box, whose magnitude is much smaller than the other
two components.
The decomposition is calculated exactly in Fourier space.
We note with the above decomposition, locally
\begin{equation}
{\vec v}_{\rm div}\cdot{\vec v}_{\rm curl} \ne 0
~~~{\rm so}~~~
\frac{1}{2} v^2 \ne
\frac{1}{2} \left(v_{\rm div}^2+ v_{\rm curl}^2 \right).
\label{ryu-e04}
\end{equation}
However, globally
\begin{equation}
\int_{\rm box} {\vec v}_{\rm div}\cdot{\vec v}_{\rm curl} d^3{\vec x} = 0
~~~{\rm so}~~~
\int_{\rm box} \frac{1}{2} v^2 d^3{\vec x} = \int_{\rm box}
\frac{1}{2} \left(v_{\rm div}^2+ v_{\rm curl}^2 \right) d^3{\vec x}.
\label{ryu-e05}
\end{equation}

The power spectra for the gas velocity and its curl and divergence
components at the present universe are shown in Figure \ref{ryu-f2}.
At long wavelengths, the amplitude of perturbations are small, so
that linear theory applies.
That is, $P_{\rm curl}(k) \rightarrow 0$  as $k \rightarrow 0$,
while $P_{\rm div}(k)$ follows the analytic theory expectation,
$P_{\rm div}(k) \sim k^{-1}$.
For wavelengths smaller than a few Mpc, nonlinearities dominate,
and we see $P_{\rm curl}(k)\ \ga\ P_{\rm div}(k)$.
$P_{\rm curl}(k)$ peaks at $\sim 5\ h^{-1}$ Mpc, and for $k$ somewhat
larger than the peak wavenumber, the spectrum follows a power law of
$k^{-5/3}$, the Kolmogorov spectrum.
$P_{\rm curl}(k)$ has most power at $\sim 2-3 h^{-1}$ Mpc, that
indicates the typical scale of nonlinear structures in the simulation.

The vorticity in the IGM as a function of gas density and temperature
is shown in the left panel of Figure \ref{ryu-f3}.
The figure exhibits a clear trend that the vorticity is larger in hotter
regions.
It is because hotter regions are occupied with the gas that has gone
through shocks of larger speed.
The vorticity generated at shocks of larger speed should be on
average larger.
Also, there are regions of high density
$(\rho_{\rm gas} / \left<\rho_{\rm gas}\right>\ \la\ 100)$
and warm temperature ($T\ \ga\ 10^4$ K), where the vorticity is large.
These regions contain the gas that was heated to high temperature
and subsequently cooled down.

\begin{figure}
\vskip -3.2cm\hskip -0.2cm
\includegraphics[width=1.05\textwidth]{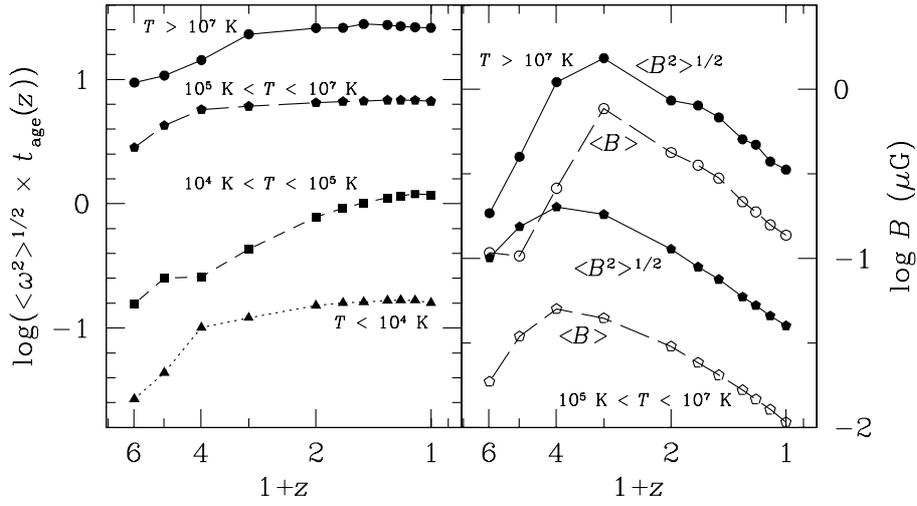}
\vskip -2.5cm
\caption{The time evolution of the rms of the vorticity for four
temperature phases of the IGM (left panel) and the time evolution of
the averaged strengths of the IGMF for the hot IGM and WHIM
(right panel) as a function of redshift $z$.
The vorticity is given in units of $t_{\rm age}(z)^{-1}$, where
$t_{\rm age}(z)$ is the age of the universe at $z$.}
\label{ryu-f4}
\end{figure}

In the left panel of Figure \ref{ryu-f4}, the root-meas-square (rms)
of the vorticity at a few redshifts for the gas in the four phases of
the IGM is shown.
Note that the vorticity shown was computed over the same comoving scales,
and normalized with the age of the universe at given $z$.
The vorticity increases as the universe evolves and the LSS of the
universe develops.
But over the period of the time presented in the figure, the vorticity,
especially in the hot IGM and WHIM, has increased just by a factor
of a few.

Figures \ref{ryu-f3} and \ref{ryu-f4} indicate that at the present epoch,
$\omega_{\rm rms} t_{\rm age} \sim$ a few $\times 10$ inside and around
clusters/groups with $T\ \ga\ 10^7$ K and $\sim 10$ in filaments which
is filled mostly with the WHIM.
On the other hand, $\omega_{\rm rms} t_{\rm age}$ is on the order
of unity in sheetlike structures and even smaller in voids.
Here, $t_{\rm age}$ is the present age of the universe, so
$\omega_{\rm rms} t_{\rm age}$ represents the number of eddy
turnovers in the age of the universe.
It takes a few turnover times for vorticity to decay and develop into
turbulence.
So it is likely that the flows in clusters/groups and filaments is
in a turbulent state, while turbulence have not significantly developed
in sheetlike structures and voids.

\begin{figure}
\vskip -4cm\hskip -2.5cm
\includegraphics[width=1.4\textwidth]{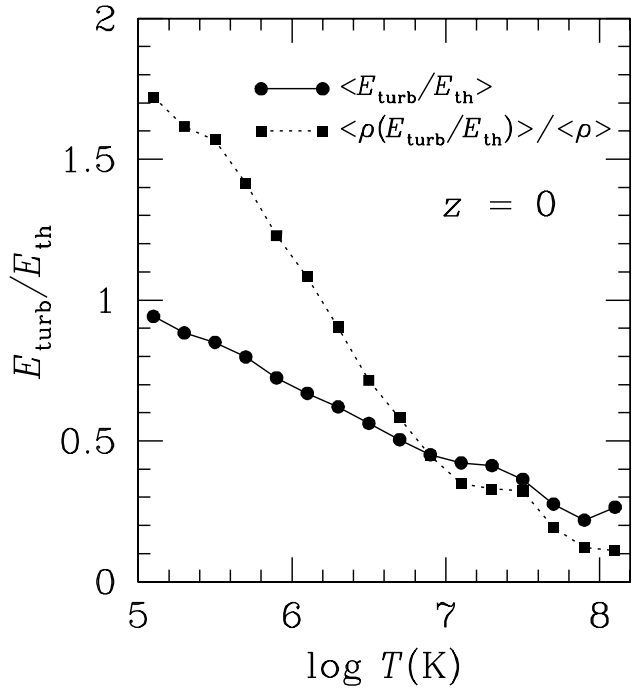}
\vskip -3cm
\caption{Turbulence to thermal energy ratio as a function of
temperature at $z=0$.
The values shown are volume-averaged and mass-averaged over
temperature bins.}
\label{ryu-f5}
\end{figure}

To estimate the energy associated with the turbulence induced in the IGM
via the cascade of the vorticity, we assume that the energy of vortical
motions, $(1/2)\rho_{\rm gas} v_{\rm curl}^2$, is transferred to that of
turbulent motions.
Then, we can regard it as the turbulence energy, $\varepsilon_{\rm turb}$.
Figure \ref{ryu-f5} shows the ratio of the turbulence to thermal
energies in clusters/groups and filaments as a function of temperature at
the present universe.
In clusters/groups with $T\ \ga\ 10^7$ K,
$\varepsilon_{\rm turb} < \varepsilon_{\rm therm}$.
Particularly the mass averaged value shows that
$\varepsilon_{\rm turb} / \varepsilon_{\rm therm}$ would be $\sim 0.1 - 0.3$
in cluster cores.
The predicted $\varepsilon_{\rm turb} / \varepsilon_{\rm therm}$ is in
a good agreement with observation \citep{sfmb04}.
Note that $M_{\rm turb} \equiv v_{\rm curl}/c_{\rm s}
= \sqrt{1.8}\ (\varepsilon_{\rm turb} / \varepsilon_{\rm therm})^{1/2}$,
where $c_{\rm s}$ is the sound speed.
Therefore, overall turbulence is subsonic in clusters/groups,
whereas it is transonic or mildly supersonic in filaments.

\subsection{Amplification of the IGMF by the turbulence in the IGM}
\label{s-igmf}

In principle, if simulations for the formation of LSS in the universe
includes magnetic fields, that is, if they are MHD, the amplification
of the seed magnetic fields by the turbulent motions in the IGM
should be able to be followed.
But, in reality, simulations with the current capacity of computing
power have too low a resolution to reproduce the full development of
turbulence inside nonlinear structures.
Also the numerical resistivity is larger than the physical resistivity
by great many order of magnitude.
As a result, the growth of magnetic fields is saturated before dynamo
action become fully operative, and the amplification of magnetic fields
can not be followed correctly, as was already pointed in \citep{kcor97}.

In order to reproduce the growth of magnetic fields by dynamo action,
a separate three-dimensional simulation of MHD turbulence in a controlled
box was performed;
incompressible, driven turbulence with initially very weak or zero
magnetic fields was simulated using a pseudospectral code \citep{cv00}.
Hyperviscosity and hyperresistivity with the Prandtl number of unity
were used.
The advantage of performing incompressible simulations using a
pseudospectral code is that the intrinsic numerical viscosity and
resistivity are virtually zero.
And by using hyperviscosity and hyperresistivity, the inertial
range can be maximized.
We point that it would take much higher resolution to achieve the
same growth rate in simulations of compressible MHD turbulence.

\begin{figure}
\vskip -2.6cm\hskip -1.cm
\includegraphics[width=1.15\textwidth]{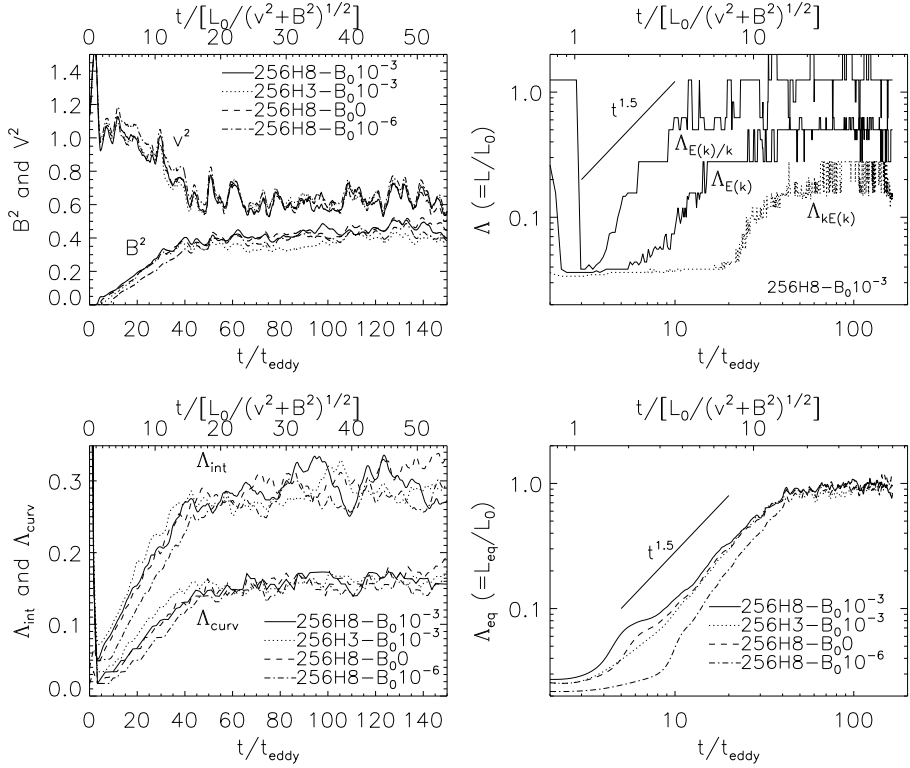}
\vskip -6.2cm
\caption{Top-left panel: The time evolution of $V^2$ and $B^2$ for
four different runs.
Here, the kinetic energy and magnetic energy densities are $V^2/2$
and $B^2/2$, respectively.
Top-right panel: Time evolution of peak scales of magnetic field
spectrum, $L_{E(k)}$, $L_{kE(k)}$, and $L_{E(k)/k}$.
Bottom-right panel: Time evolution of the energy equipartition scale,
$L_{\rm eq}$.
Bottom-left panel: Time evolution of the integral scale, $L_{\rm int}$,
and the curvature scale, $L_{\rm curv}$.
$\Lambda$'s are the scales normalized with the energy injection scale
$L_0$.
See the text for details.}
\label{ryu-f6}
\end{figure}

The top-left panel of Figure \ref{ryu-f6} shows the time evolution
of kinetic and magnetic energies for four different simulations:
256H8-B$_010^{-3}$, 256H3-B$_010^{-3}$, 256H8-B$_00$, and
256H8-B$_010^{-6}$.
Simulations are denoted with 256Y-$B_0$Z, where $256$ refers to
the number of grid points in each spatial direction, Y refers to
hyperdissipation (H) and its order, and Z refers to the strength of
the external magnetic fields.
The turbulence was driven at the scale of $L_0 \sim
(1/2)L_{\rm box}$ where $L_{\rm box}$ is the computational box
size, and the driving strength was set so that the total energy
is $E_{\rm tot} \equiv E_{\rm kin} + E_B \sim 1$ at saturation.
The time is given in units of the eddy turnover time that is
defined as the inverse of vorticity at driving scale,
$t_{\rm eddy} \equiv 1/\omega_{\rm driving}$, at saturation.
See \citet{cvbl09} for details of simulations.

The amplification of magnetic fields by turbulence dynamo is shown
in the figure.
It does not sensitively depend on the initial magnetic field
strength once it is sufficiently weak as well as details of
simulations including the dissipation prescription.
The evolution of magnetic fields goes through three stages:
the initially exponential growth when the back reaction of magnetic
fields is negligible, then the linear growth when the back reaction
starts to operate, and the final saturation.
By fitting the evolution, we model the growth and saturation of
magnetic field energy as
\begin{eqnarray}
E_B = \left\{ \begin{array}{ll}
0.04 \times \exp\left[(t/t_{\rm eddy}-4)/0.36\right]
&{\rm for}\ t/t_{\rm eddy}<4 \\
(0.36/41) \times (t/t_{\rm eddy}-4) + 0.04
&{\rm for}\ 4<t/t_{\rm eddy}<45 \\
0.4 &{\rm for}\ t/t_{\rm eddy}>45
\end{array} \right.
\label{ryu-e06}
\end{eqnarray}

Along with the amplification of magnetic field strength, the
magnetic fields become coherent through the inverse cascade.
For the quantification, different characteristic lengths of
magnetic fields can be defined in MHD turbulence:
the peak scale of the spectrum of magnetic fields, $L_{E(k)}$,
the scale containing the largest energy of magnetic fields, $L_{kE(k)}$,
the peak scale of the spectrum of projected magnetic fields,
$L_{E(k)/k}$, the energy equipartition scale,
$L_{\rm eq} (= 2\pi/k_{\rm eq})$, defined as
\begin{equation}
\int_{k_{\rm eq}}^{k_{\rm max}} E_v(k)\ dk =
\int_0^{k_{\rm max}} E_b(k)\ dk,
\label{ryu-e07}
\end{equation}
the integral scale, $L_{\rm int}$, defined as
\begin{equation}
L_{\rm int} = 2\pi \frac{\int E_b(k)/k\ dk }{\int E_b(k)\ dk},
\label{ryu-e08}
\end{equation}
and the curvature scale, $L_{\rm curv}$, defined as a typical radius of
curvature of field lines.

The rest of Figure \ref{ryu-f6} show the time evolution of characteristic
lengths of magnetic fields.
At saturation, the peak of magnetic field spectrum, $L_{E(k)}$, occurs at
$\sim L_0/2$, where $L_0$ is the energy injection scale, while the most
energy containing scale, $L_{kE(k)}$ is $\sim L_0/5$.
During the stage of magnetic field amplification, the energy
equipartition scale, $L_{\rm eq}$, shows a power-law increase of
$\sim t^{1.5}$, while the integral scale, $L_{\rm int}$, and the
curvature scale, $L_{\rm curv}$, show a linear increase.
The equipartition, integral, and curvature scales saturate at $\sim L_0$,
$\sim 0.3L_0$, and $\sim 0.15L_0$, respectively.
See \citet{cr09} for further details of characteristic lengths in
MHD turbulence with very weak or zero mean magnetic fields.

The results of the incompressible MHD turbulence simulation were
convoluted to the data of the stimulation for the formation of the
LSS of the universe to get the IGMF.
For the estimation of the magnetic field strength, it was assumed
that a fraction of the turbulence energy is converted into the
magnetic energy.
The fraction was expressed as a function of the number of local
eddy turns over the age of the universe, so
\begin{equation}
\varepsilon_B = \varepsilon_{\rm curl} \cdot
\phi(\omega \times t_{\rm age}).
\label{ryu-e09}
\end{equation}
For the fraction $\phi\left(\omega \times t_{\rm age}\right)$,
the fitting formula in Equation (\ref{ryu-e06}) was used.
Then, the magnetic field strength was calculated as
$(8\pi \varepsilon_B)^{1/2}$.

The resulting magnetic field strength as a function of gas density
and temperature is shown in the right panel of Figure \ref{ryu-f3}.
On average, the IGMF is predicted to be stronger in hotter and denser
regions.
It is because with the turbulence energy is larger in hotter and
denser regions.
Also the conversion factor, $\phi$, is larger in hotter and denser
regions.

In the right panel of Figure \ref{ryu-f4}, the volume-averaged and
rms values of the magnetic field strength at a few redshifts for
the gas in the hot IGM and WHIM are shown.
Our scenario predicts that the magnetic field strength would be
$\left< B \right>\ \ga\ 1\ \mu$G inside clusters/groups,
$\sim 0.1\ \mu$G around clusters/groups, and $\sim 10$ nG in
filaments at the present universe.
The magnetic fields should be much weaker in sheetlike structures
and voids.
But as noted in Section \ref{s-turb}, turbulence is not fully developed
there.
So our estimation of the IGMF in those regions should not be applicable.

In each temperature range, the magnetic fields were stronger in the past
at $z \sim$ a few. Our calculation indicates that the vorticity has not
changed much since $z \sim$ a few (see the left panel of Figure \ref{ryu-f4}),
nor the vortical component of flow velocity \citep{rk08}.
So the stronger magnetic fields in the past should be due to the higher
density.  We point that clusters were virialized around $z \sim 1$ or so,
so the density within the virial radius of $\sim 1$ Mpc has not changed
much since then.  But our estimation of the magnetic field strength
for the hot IGM with $T > 10^7$ K also includes the gas in cluster
outskirt, which extends far beyond the virial radius, up to several
Mpc or even larger \citep[see, e.g.][]{rkhj03};
the density of the gas there was higher in the past.
We also note that the magnetic fields, when averaged all over
the computational box, were weaker in the past, because the fraction
of the strong field regions was smaller.

Based on a kinetic theory, assuming the Kolmogorov spectrum for
turbulence flow motions, \citet{kcor97} and \citet{kz07} estimated
that the magnetic field strength in clusters would be a few $\mu$G.
Our result agrees with the previous work.
It also matches well with the observed strength of magnetic fields
in the ICM, which is discussed in Section \ref{s-intr}.
On the other hand, our prediction of $B \sim 10$ nG in filaments
is within but lower than the upper limit of $\sim 0.1 \mu$G which is
imposed from RMs outside of clusters \citep{rkb98,xkhd06}.

The characteristic lengths of the IGMF in our scenario can be
conjectured.
In clusters, turbulence is near the saturation stage with
$t/t_{eddy} \sim 30$ (see Figures \ref{ryu-f4} and \ref{ryu-f6}).
Then, for instance, we may estimate that the peak scale of magnetic
field spectrum and the integral scale would $L_{E(k)}/L_0 \sim 0.4$
and $L_{\rm int}/L_0 \sim 0.2$.
If we take the energy injection scale $L_0 \sim 100$ kpc, which
is approximately the scale height of cluster core,
$L_{E(k)} \sim 40$ kpc and $L_{\rm int} \sim 20$ kpc, respectively.
In filaments, on the other hand, with $t/t_{eddy} \sim 10$, turbulence
is expected to be still in the linear growth stage, and
$L_{E(k)}/L_0 \sim 1/10$ and $L_{\rm int}/L_0 \sim 1/15$.
We may take the energy injection scale $L_0 \sim 5$ Mpc, which is the
typical thickness of filaments.
Then, $L_{E(k)} \sim 0.5$ Mpc and $L_{\rm int} \sim 0.3$ Mpc, respectively.

\subsection{Contribution from AGNs}\label{s-agns}

There is a possibility that the IGMF is further strengthened by the
magnetic fields ejected through jets from back holes in AGNs
\citep[see, e.g.,][]{kdlc01}.
Strong magnetic fields almost certainly arise in accretion disks
surrounding black holes.
These fields may find their way into the IGM via magnetically dominated
jets.
The potential field strength due to this process can be estimated
as follows \citep[see, e.g.,][]{hoyle69}.
The rotational energy associated with the central compact objects of
mass $M$ which power AGNs can be parametrized as $fMc^2$ where $f < 1$.
If we assume equipartition between rotational and magnetic energies
within a central volume $V_c$, we find
\begin{equation}
B_c \sim \left (\frac{8\pi fMc^2}{V_c}\right )^{1/2}.
\label{agnb}
\end{equation}
If this field then expands adiabatically to fill a volume $V$ in the IGM,
one finds $B \sim B_c\left (V_c/V\right )^{2/3}$.
Considering the values of $M=10^9\,M_\odot$, $f=0.1$, and
$V \simeq \left (1\ {\rm Mpc}\right)^3$, we get $B \sim 10$ nG.
Note this field strength is comparable to that estimated in filaments
with turbulence dynamo.
So this process may enhance the strength of the IGMF by a factor of two
or so.

Although there are some RM observations that indicate strong magnetic
fields in some of jets, however, it is not yet clear whether all jets are
magnetically dominated.
Also the details of the population of AGN jets and the volume filling
fraction of magnetic fields in the universe by this process need to
be further worked out.

\section{Astrophysical Implications of the IGMF}\label{s-impl}

The existence of the IGMF, especially in filaments, can have
noticeable implications on a variety of astrophysical phenomena.
In this section, we discuss two: the effect on the propagation of
ultra-high energy cosmic rays (UHECRs) and the inducement
of Faraday rotation.

\subsection{Propagation of ultra-high energy cosmic rays}

UHECRs are known to originate from extragalactic sources.
Hence, on their path through the intergalactic space, the trajectories
of UHECRs are deflected by the magnetic fields between sources and us,
the IGMF as well as the galactic magnetic field (GMF).
\citet{dkrc08} studied the effect of the IGMF described in
Section \ref{s-ampl} on the propagation of UHECRs.
Under the premise that the sources of UHECRs are strongly associated
with the LSS of the universe, super-GZK protons of $E \geq 10^{19}$ eV
were injected by AGN-like sources located inside clusters of galaxies.
Then, the trajectories of the protons were followed, while taking
account of the energy loss due to interactions with the cosmic
microwave background (CMB) radiation.
\citet{dkrc08} found that the deflection of UHECR trajectories is
caused mostly by the IGMF in filaments, rather than the the intracluster
magnetic field in clusters;
it is because filaments fill a larger fraction of volume,
although their magnetic fields are weaker, than clusters.
With the gyroradius of protons
\begin{equation}
r_g = 1\ {\rm Mpc} \left(\frac{E_{\rm UHECR}}{10^{19}\ {\rm eV}}\right)
\left(\frac{B}{10\ {\rm nG}}\right)^{-1},
\label{ryu-e10}
\end{equation}
that is, with the gyroradius corresponding typical filament size
for the magnetic fields typical in filaments, the deflection due to
the IGMF in filaments is expected to be significant.
Indeed, the deflection angle between the arrival direction of super-GZK
protons and the sky position of their actual sources was found to be
quite large with the mean value of $\langle \theta \rangle \sim 15^{\circ}$
and the median value of $\tilde \theta \sim 7 - 10^{\circ}$.

The above deflection is much larger than the deflection by the GMF;
the deflection angle due to the GMF was predicted to be a few degree
\citep[see, e.g.,][]{ts08}.
As a matter of fact, the deflection angle of
$\langle \theta \rangle \sim 15^{\circ}$ is also much than the
angular window of $3.1^{\circ}$ used by the Auger collaboration
in the study of the correlation between their highest energy UHECR
events and nearby AGNs \citep{auger07a}.
Although the deflection angle is large, \citet{rdk10} noticed that
in the work of \citet{dkrc08}, the separation angle between the arrival
direction of super-GZK protons and the sky position of nearest AGNs is
substantially smaller with
$\langle S \rangle \sim 3.5 - 4^{\circ}$, which is similar to the mean
angular distance in the sky to nearest neighbors among AGNs.
This mean separation angle is comparable to the angle used in
the correlation study by the Auger collaboration.
This is a direct consequence of the fact that the sources and us, as well
as the IGMF, all trace the matter distribution of the universe.
That is, although the IGMF described in Section \ref{s-ampl} predicts
larger deflection of UHECRs, it is not necessarily inconsistent with
the intervening magnetic fields implied in the Auger experiment.

\subsection{Faraday rotation induced by the IGMF}

The IGMF described in Section \ref{s-ampl} induces RM.
With characteristic lengths smaller than the dimension of clusters
or filaments, the inducement of RM is a random walk process;
the standard deviation of RM is
\begin{equation}
\sigma_{\rm RM} = 0.81\ {\bar n_e}\ B_{\parallel {\rm rms}}
\ \sqrt{\left(\frac{3L_{\rm int}}{4}\right)L}~{\rm rad~m^{-2}},
\label{ryu-e11}
\end{equation}
where $n_e$, $B_{\parallel {\rm rms}}$, and $L$ are in units of cm$^{-3}$,
$\mu$G, and pc, respectively \citep{cr09}.
$B_{\parallel {\rm rms}}$ is the rms strength of line-of-sight magnetic
field and $L$ is the path length.
Here, the coherence length for RM is given as $l = (3/4)L_{\rm int}$.
For clusters, with ${\bar n_e} \sim 10^{-3}$ cm$^{-3}$, $B_{\rm rms} \sim$
a few $\mu$G, $L \sim 1$ Mpc, and $L_{\rm int} \sim 20$ kpc (see Section
\ref{s-ampl}), we get $\sigma_{\rm RM} \sim 100$ rad m$^{-2}$, which agrees
with the observed RM in clusters \citep{ckb01}.
The magnetic field strength in filaments quoted in Section \ref{s-ampl}
is $\langle B\rangle \sim 10$ nG.
But the value depends on how it is averaged; that is,
$\langle B^2\rangle^{1/2} \sim$ a few $\times~10$ nG,
$\langle \rho B\rangle/\langle \rho\rangle \sim 0.1~\mu$G,
and $\langle (\rho B)^2\rangle^{1/2}/\langle \rho^2\rangle^{1/2} \sim$
a few $\times~0.1~\mu$G, in the warm-hot intergalactic medium
(WHIM) with $T = 10^5 - 10^7$ K which mostly composes filaments.
The average value of $\langle (\rho B)^2\rangle^{1/2}/\langle
\rho^2\rangle^{1/2}$ should be relevant to RM.
With ${\bar n_e} \sim 10^{-5}$ cm$^{-3}$, $B_{\rm rms} \sim 0.3\ \mu$G,
$L \sim 5$ Mpc, and $L_{\rm int} \sim 300$ kpc, we get
$\sigma_{\rm RM} \sim 1$ rad m$^{-2}$ through filaments.

\begin{figure}
\vskip 0cm\hskip 1.5cm
\includegraphics[width=0.8\textwidth]{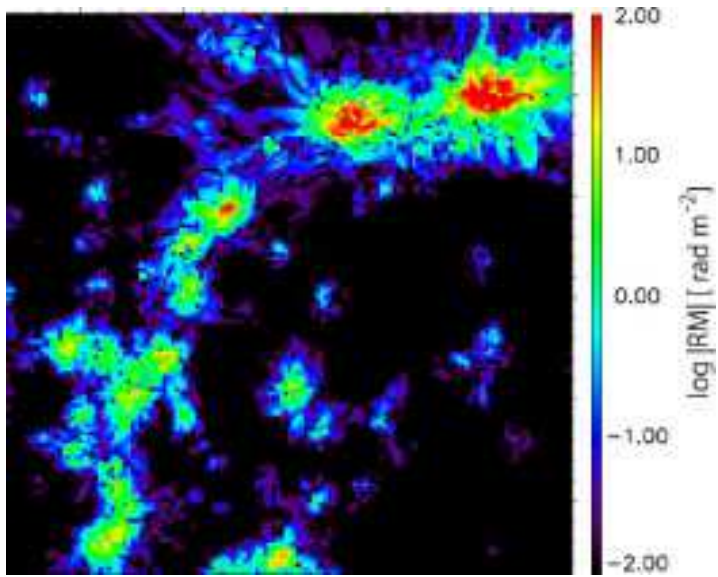}
\vskip 0.2cm
\caption{RM map of path length of $L=100\ h^{-1}$ in the local universe
o $(28\ h^{-1}{\rm Mpc})^2$ area at $z=0$.}
\label{ryu-f7}
\end{figure}

\begin{figure}
\vskip 0cm\hskip 0.7cm
\includegraphics[width=0.8\textwidth]{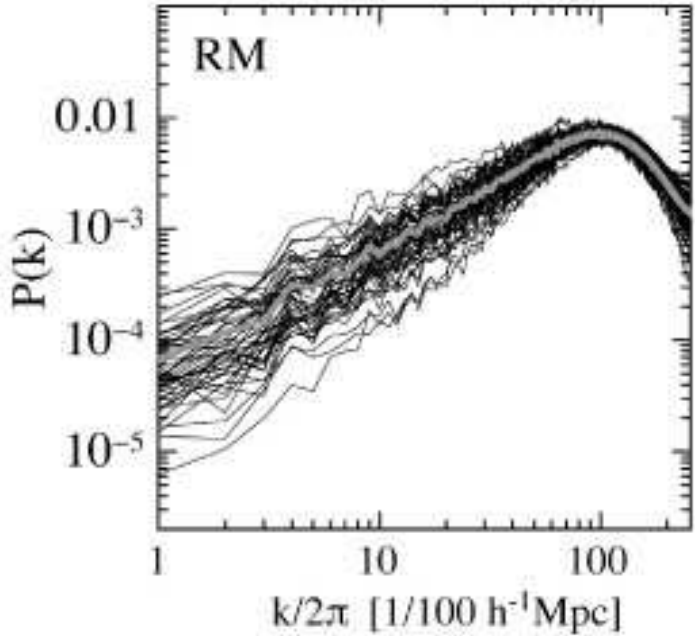}
\vskip 0.2cm
\caption{Two-dimensional power spectra of RM for $3\times 16$
two-dimensional projected maps from 16 simulations with different
realizations of initial condition to compensate cosmic variance.
The thick gray line shows the average.}
\label{ryu-f8}
\end{figure}

\citet{ar10} studied in details the RM due to the magnetic fields in
filaments using the IGMF described in Section \ref{s-ampl}.
Figure \ref{ryu-f7} shows a typical RM map in $(28\ h^{-1}{\rm Mpc})^2$
area; the spatial distribution of RM traces the large-scale distribution
of matter, showing two clusters and a filamentary structure containing
several groups.
The resultant RM is dominantly contributed by the density peak along
line of sight.
The rms of RM through filaments at the present universe was
predicted to be $\sim 1\ {\rm rad\ m^{-2}}$, which agrees with the
estimation above \citep{cr09}.
Figure \ref{ryu-f8} shows the two-dimensional power spectrum of the RM
in the local universe within $100\ h^{-1}$ Mpc;
$P_{\rm RM}(k) \sim |{\rm RM}(\vec{k})|^2 k$, where ${\rm RM}(\vec{k})$
is the Fourier transform of ${\rm RM}(\vec{x})$.
The power spectrum of RM peaks at $k\sim 100$, which corresponds to
$\sim 1\ h^{-1}$ Mpc.
In addition, \citet{ar10} predicted that the probability distribution
function (PDF) of $|{\rm RM}|$ through filaments follows the log-normal
distribution.
We note that RM $\sim 1\ {\rm rad\ m^{-2}}$ is an order of magnitude
smaller than the values of $|{\rm RM}|$ toward the Hercules and
Perseus-Pisces superclusters reported in \citet{xkhd06}.
The difference is mostly due to the mass-weighted path length;
the value quoted by \citet{xkhd06} is about two orders of magnitude
larger than ours.

RM $\sim 1\ {\rm rad\ m^{-2}}$ due to the IGMF in filaments is
too small to be confidently observed with currently available
facilities.
In addition, the galactic foreground of $\sim 10\ {\rm rad\ m^{-2}}$
(toward halo) poses an additional challenge for its observation.
The next generation radio interferometers, however, are expected to
be able to observe the RM.
Particularly, the SKA could measure RM for $\sim 10^8$ polarized
extragalactic sources across the sky with an average spacing
of $\sim 60$ arcsec between lines of sight
\citep[papers in][]{cr04} \citep{kra09}, enabling
us to investigate the IGMF in the LSS of the universe.

\section{Structure Formation and Magnetic Fields}\label{s-sfmg}

The existence of magnetic fields in the early universe, although
probably weak, can have consequences on the structure formation
itself.  In this section, we briefly review recent developments
in our understanding of magnetized structure formation.

\subsection{The linear regime}\label{s-line}

Studies of magnetized structure formation go back to the late 1960s 
with the early efforts based on Newtonian gravity and the relativistic 
approaches being a relatively recent addition. All treatments typically 
work within the ideal MHD approximation and
look at the effects of the magnetic Lorentz force on density 
inhomogeneities \citep{rr71,wa78,fe80,kor96,bfj97,tb97,tm00a}. 
These effects generally come in the form of scalar, vector and 
(trace-free) tensor distortions. The former are those commonly 
referred to as density perturbations and represent over-densities 
or under-densities in the matter distribution. Vector inhomogeneities 
describe rotational (i.e. vortex-like) density perturbations. 
Finally, tensor-type density inhomogeneities correspond to shape 
distortions.\footnote{It should be made clear that trace-free 
tensor inhomogeneities (i.e. shape deformations) and pure-tensor 
distortions (i.e. gravitational waves) are two different 
types of perturbations.} Following \citep{tb97,bmt07}, 
we define the scalar
\begin{equation}
\Delta\equiv  {a^2\over\rho}\,{\rm D}^2\rho\,,
\label{divs}
\end{equation}
which describes linear density perturbations and corresponds 
to the more familiar density contrast $\delta\rho/\rho$.  
Note that positive values for $\Delta$ indicate over-densities 
and negative ones under-densities. Also, $a$ is the cosmological 
scale factor and ${\rm D}^2={\rm D}^a{\rm D}_a$ is the 3-D 
Laplacian operator that
corresponds to an observer moving with 4-velocity 
$u_a$.\footnote{For an observer with 4-velocity $u_a$ 
(so that $u_au^a=-1$), the tensor $h_{ab}=g_{ab}+u_au_b$ 
projects orthogonal to $u_a$ and ${\rm D}_a=h_a{}^b\nabla_b$ 
defines the covariant derivative operator of the spatial 
hypersurfaces ($g_{ab}$ is the space-time metric and 
$\nabla_a$ the associated covariant derivative).} 
In a perturbed, weakly magnetized and spatially flat
Friedmann-Robertson-Walker (FRW) universe, 
the above defined scalar
evolves according to
\begin{equation}
\dot{\Delta}= 3wH\Delta- (1+w)\mathcal{Z}+ {3\over2}\,
c_{\rm a}^2(1+w)H\mathcal{B}\,,
\label{dotDelta}
\end{equation}
where over-dots denote proper-time derivatives (relative to the
$u_a$-frame). Also, $w=p/\rho$ is the (constant)
barotropic index of the matter, $H=\dot{a}/a$ is the 
background Hubble parameter (with $\Theta=3H$ there) and 
$c_{\rm a}^2=B^2/\rho(1+w)$ is the square of the Alfv\'en 
speed. The variables $\mathcal{Z}=a^2{\rm D}^2\Theta$ 
and $\mathcal{B}= (a^2/B^2){\rm D}^2 B^2$ describe linear 
inhomogeneities in the smooth Hubble expansion and in 
the magnetic energy density respectively. Then, to first order,
\begin{eqnarray}
\dot{\mathcal{Z}}&=& -2H\mathcal{Z}- {1\over2}\,\rho\Delta+
{1\over4}\,c_{\rm a}^2(1+w)\rho\mathcal{B}- {c_s^2\over1+w}\,
{\rm D}^2\Delta- {1\over2}\,c_{\rm a}^2{\rm D}^2\mathcal{B}
\label{dotcZ}
\end{eqnarray}
and
\begin{equation}
\dot{\mathcal{B}}= {4\over3(1+w)}\,\dot{\Delta}\,,
\label{dotcB}
\end{equation}
respectively. Note that $c_s^2=\dot{p}/\dot{\rho}$ is the square 
of the adiabatic sound speed and we have assumed that 
$B^2\ll\rho$, given the relative weakness of the magnetic fields.

Equation (\ref{dotDelta}) shows that magnetic fields are 
generic sources of linear density perturbations. Indeed, 
even if $\Delta$ and $\mathcal{Z}$ are zero initially, 
$\dot{\Delta}$ will take nonzero values solely due to the 
magnetic presence. Also, Equation (\ref{dotcB}) ensures 
that perturbations in the magnetic field energy density 
evolve in tune with their matter counterparts 
(i.e. $\mathcal{B}\propto\Delta$). Finally, we should 
emphasize that only the pressure part of the Lorentz 
force contributes to the linear relations (\ref{dotDelta}) 
and (\ref{dotcZ}).\footnote{The Lorentz force splits as 
$\varepsilon_{abc}B^b{\rm curl}B^c
={\rm D}_a/2B^2- B^b{\rm D}_bB_a$, 
with the former term corresponding to the magnetic pressure 
and the latter to the field's tension. The effects of the 
$B$-field on (scalar) density perturbations propagate via 
the divergence of the Lorentz force. To leading order, 
the latter is given by 
${\rm D}^a (\varepsilon_{abc}B^b{\rm curl}B^c)
={\rm D}^2B^2/2-B^b{\rm D}_b{\rm D}^aB_a-KB^2/a^2$, 
where $K=0,\pm1$ represents the background 3-curvature index. 
Given that ${\rm D}^aB_a=0$ at the ideal-MHD limit, 
the magnetic tension effects are not included in this 
perturbative level, unless the spatial curvature of the 
FRW model is accounted for \citep{tm00a,tm00b}.}

The system of Equations (\ref{dotDelta}) -- (\ref{dotcB}) has
analytical solutions in the radiation and dust eras \citep{tb97}.
Before equipartition, when $w=1/3=c_s^2$, $H=1/2t$, $\rho=3/4t^2$
and $c_{\rm a}^2=3B^2/4\rho=\,$constant, large-scale magnetized
density perturbations obey a power-law solution. In particular,
on super-horizon scales and keeping only the dominant growing 
and decaying modes, one arrives at \citep{tm00a,bmt07}
\begin{equation}
\Delta= \mathcal{C}_1t^{-1/2+10c_{\rm a}^2/9}+
\mathcal{C}_2t^{1-4c_{\rm a}^2/9}\,.
\label{lsrDelta}
\end{equation}
In the absence of magnetic fields, we recover the standard growing
and decaying modes of $\Delta\propto t$ and $\Delta\propto t^{-1/2}$ 
respectively. So, the magnetic presence has reduced the growth rate
of the density contrast by $4c_{\rm a}^2/9$.

Well inside the horizon we can no longer ignore the role of 
the pressure gradients. There, the $k$-mode oscillates like 
a magneto-sonic wave with
\begin{equation}
\Delta_{(k)}\propto \sin\left[c_s\left(1+{2\over3}\,c_{\rm a}^2\right)
\left({\lambda_H\over\lambda_k}\right)_0\sqrt{t\over t_0}\, \right]\,,
\label{ssrDelta}
\end{equation}
where $\lambda_k=a/k$ is the perturbed scale and $\lambda_H=1/H$ 
the Hubble horizon \citep{tm00a,bmt07}. Here, the magnetic 
pressure increases the effective sound speed and therefore the 
oscillation frequency. The former makes the Jeans length larger 
than in non-magnetized models. The latter brings the peaks of 
short-wavelength oscillations in the radiation density closer, 
leaving a potentially observable signature in the CMB spectrum 
\citep{adgr96}.

When dust dominates, $w=0=c_s^2$, $H=2/3t$, $\rho=4/3t^2$ and
$c_{\rm a}^2=B^2/\rho\propto t^{-2/3}$. Then, on superhorizon
scales, the main growing and decaying modes of the density contrast
are \citep{tb97,bmt07}
\begin{equation}
\Delta= \mathcal{C}_1t^{\alpha_1}+ \mathcal{C}_2t^{\alpha_2}\,,
\label{dDelta}
\end{equation}
with $\alpha_{1,2}= -[1\pm5\sqrt{1-(32/75) (c_{\rm a}\,\lambda_H
/\lambda_k)_0^2}]/6$. In the absence of magnetic fields we recover
again the standard solution with $\alpha_1=2/3$ and $\alpha_2=-1$.
Thus, as with the radiation era before, the magnetic presence 
slows down the growth rate of density perturbations. Also, since
$\mathcal{B}\propto\Delta$ [see Equation (\ref{dotcB})], the above
describes the linear evolution of the magnetic energy-density
perturbations as well. This means that cosmological magnetic fields
trapped inside an overdense region of the post-recombination 
universe could grow by approximately two to three orders of magnitude. 
Note that the aforementioned increase is different from the 
one occurring during the subsequent, nonlinear contraction 
of a protogalactic cloud (see Section \ref{s-nonl} below).

The field pressure also leads to a magnetically induced Jeans 
length, below which density perturbations cannot grow 
\citep{sb98,ss05}. As a fraction of the Hubble radius, 
this purely magnetic Jeans scale is \citep{tm00a,bmt07}
\begin{equation}
\lambda_J\sim c_{\rm a}\lambda_H\,.
\label{mJeans}
\end{equation}
Setting $B\sim10^{-9}$ G, which is the maximum homogeneous field
strength allowed by the CMB \citep{zeld70a,bfs97}, we find that
$\lambda_J\sim10$ kpc. Alternative, magnetic fields close to
$10^{-7}$ G, like those found in galaxies and galaxy clusters,
give $\lambda_J\sim1$ Mpc. The latter lies intriguingly close
to the size of a cluster of galaxies.

Overall, the magnetic effect on density perturbations is rather
negative. Although magnetic fields generate this type of distortions,
they do not help them to grow. Instead, the magnetic presence either
suppresses the growth rate of density perturbations, or increases
the effective Jeans length and therefore the domain where these
inhomogeneities cannot grow. This negative role of magnetic fields,
which was also observed in the Newtonian treatment of \citep{rr71},
reflects the fact that only the pressure part of the Lorentz force
has been incorporated into the equations. When the tension component
(i.e. the elasticity of the field lines) is also accounted for, the
overall magnetic effect can change and in some cases it could even
reverse \citep{tm00b}.

Magnetic fields also induce and affect rotational, vortex-like,
density inhomogeneities \citep{wa78,tm00a}. To linear
order, these are described by the vector
$\mathcal{W}_a= -(a^2/2\rho)\varepsilon_{abc}{\rm D}^b{\rm D}^c\rho$,
with $\varepsilon_{abc}$ representing the 3-D Levi-Civita tensor.
Then, on an spatially flat FRW background,
\begin{equation}
\ddot{\mathcal{W}}_a= -4H\dot{\mathcal{W}}_a-
{1\over2}\,\rho\mathcal{W}_a+
{1\over3}\,c_{\rm a}^2{\rm D}^2\mathcal{W}_a\,,
\label{ddotW}
\end{equation}
after matter-radiation equality \citep{tm00a,bmt07}. Defining
$\lambda_{\rm a}=c_{\rm a}\lambda_H$ as the Alfv\'en horizon,
we may write the associated solution in the form
\begin{equation}
\mathcal{W}_{(k)}= \mathcal{C}_1t^{\alpha_1}+
\mathcal{C}_2t^{\alpha_2}\,,
\label{dW}
\end{equation}
with $\alpha_{1,2}=
-[5\pm\sqrt{1-(48/9)(\lambda_{\rm a}/ \lambda_k)^2_0}]/6$.
On scales far exceeding the Alfv\'en horizon, 
$\lambda_{\rm a}\ll\lambda_k$
and the perturbed mode decays as $\mathcal{W}\propto t^{-2/3}$.
This rate is considerably slower than $\mathcal{W}\propto t^{-1}$,
the decay rate associated with magnetic-free dust cosmologies.
Well inside $\lambda_{\rm a}$, on the other hand, magnetized
vortices oscillate like Alfv\'en waves, with \citep{tm00a}
\begin{equation}
\mathcal{W}_{(k)}\propto t^{-5/6}\cos\left[{2\sqrt{3}\over9}
\left({\lambda_{\rm a}\over\lambda_k}\right)_0\ln t\right]\,.
\label{sscW}
\end{equation}
Thus, the effect of magnetic fields on a given vortex mode is
to reduce its standard depletion rate. Analogous is the magnetic
effect on $\omega_a$, namely on the vorticity proper. 
Hence, magnetized cosmologies appear to rotate faster than 
their magnetic-free counterparts. In contrast to density 
perturbations, magnetic fields seem to favor the presence 
of vorticity. This qualitative difference should probably 
be attributed to the fact that the tension part of the Lorentz 
force also contributes to Equation (\ref{ddotW}).

In addition to scalar and vector perturbations, magnetic fields
also generate and affect tensor-type inhomogeneities that 
describe shape-distortions in the density distribution 
\citep{tm00a}. An initially spherically symmetric inhomogeneity, 
for example, will change shape due to the magnetically induced 
anisotropy. All these effects result from the Lorentz force. 
Even when the latter is removed from the system, however, 
magnetic fields remain active. Due to its energy density and 
anisotropic nature, for example, magnetism affects both 
the local and long-range gravitational fields. The anisotropic 
magnetic pressure, in particular, leads to shear distortions 
and subsequently to gravitational-wave production
\citep{cd02,ts02,wa10}. Overall, magnetic fields are a very 
versatile source. They are also rather unique in nature, 
since they are the only known vector source of energy. 
An additional unique magnetic feature, which remains relatively 
unexplored, is the field's tension. When we add to all these 
the widespread presence of magnetic fields in the universe, 
it is not unreasonable to say that
no realistic structure formation scenario should a priori
exclude them.

\subsection{Aspects of the nonlinear regime}\label{s-nonl}

The evolution of large-scale magnetic fields during the nonlinear
stages of structure formation is addressed primarily by means of
numerical simulations. The reason is the high complexity of the
nonlinear MHD equations, which considerably hampers analytical
studies, unless certain simplifying assumptions are imposed.

The simplest approximation is to assume spherically symmetric
compression. Realistic collapse, however, is not isotropic.
In fact, when magnetic fields are present, their generically
anisotropic nature makes the need to go beyond spherical
symmetry greater. Certain aspects of anisotropic contraction
can be analytically studied within the Zeldovich approximation
\citep{zeld70b}. The latter is based on a simple ansatz, which
extrapolates to the nonlinear regime a well known linear result.
The assumption is that the irrotational and acceleration-free
linear motion of dust, also holds during the early nonlinear
stages of galaxy formation. This allows the analytical treatment
of the nonlinear equations, leading to solutions that describe
anisotropic (one dimensional) collapse and to the formation of
the well-known Zeldovich pancakes.

Suppose that magnetic fields are frozen into a highly conductive
protogalactic cloud that is falling into the (Newtonian) potential
wells formed by the Cold Dark Matter (CDM) sector.
\footnote{The Newtonian theory is a very good approximation,
since we are dealing with non-relativistic matter and the scales
of interest are well inside the curvature radius of the universe.}
Relative to the physical coordinate system $\{r^{\alpha}\}$,
the motion of the fluid velocity is $u_{\alpha}=3Hr_{\alpha}+v_{\alpha}$,
where $H=\dot{a}/a$ is the Hubble parameter of the unperturbed
FRW background and $v_{\alpha}$ is the peculiar velocity of
the fluid (with $\alpha=1,2,3$). Then, the magnetic field induction
equation reads \citep{bmt03}
\begin{equation}
\dot{B}_{\alpha}= -2HB_{\alpha}- {2\over3}\,\vartheta B_{\alpha}+
\sigma_{\alpha\beta}B^{\beta}\,,
\label{ZdotB}
\end{equation}
where $\vartheta=\partial^{\alpha}v_{\alpha}$ and
$\sigma_{\alpha\beta}=\partial_{\langle\beta}v_{\alpha\rangle}$
are the peculiar volume scalar and the peculiar shear tensor
respectively.\footnote{When dealing with purely baryonic collapse,
the Zeldovich ansatz only holds during the early stages of the
nonlinear regime, when the effects of the fluid pressure are
negligible. Assuming that the contraction is driven by non-baryonic
CDM, means that we can (in principle) extend the domain of
the Zeldovich approximation further.} The former takes negative
values (i.e. $\vartheta<0$), since we are dealing with a
protogalactic cloud that has started to turn around and collapse.
Also note that the first term in the right-hand side of Equation
(\ref{ZdotB}) reflects the background expansion, the second is
due to the peculiar contraction and the last carries the
anisotropic effects. Introducing the rescaled magnetic field
$\mathcal{B}_{\alpha}= a^2B_{\alpha}$, the above expression
recasts into
\begin{equation}
\mathcal{B}^{\prime}_{\alpha}= -{2\over3}\,\tilde{\vartheta}
\mathcal{B}_{\alpha}+ \tilde{\sigma}_{\alpha\beta}
\mathcal{B}^{\beta}\,,
\label{ZrB}
\end{equation}
with primes indicating differentiation with respect to the scale
factor. Also $\vartheta=aH\tilde{\vartheta}$ and
$\sigma_{\alpha\beta}=aH\tilde{\sigma}_{\alpha\beta}$,
where $\tilde{\vartheta}=\partial^{\alpha}\tilde{v}_{\alpha}$ and
$\tilde{\sigma}_{\alpha\beta}= \partial_{\langle\beta}
\tilde{v}_{\alpha\rangle}$ (with $\tilde{v}_{\alpha}=
ax^{\prime}_{\alpha}$ and $v_{\alpha}=aH\tilde{v}_{\alpha}$).
In the shear eigen-frame, $\tilde{\sigma}_{\alpha\beta}=
(\tilde{\sigma}_{11},\, \tilde{\sigma}_{22},\,\tilde{\sigma}_{33})$
and Equation (\ref{ZrB}) leads to
\begin{equation}
\mathcal{B}^{\prime}_1= -{2\over3}\,\tilde{\vartheta}
\mathcal{B}_1+ \tilde{\sigma}_{11}\mathcal{B}_1\,,
\label{ZrB1}
\end{equation}
with exactly analogous relations for the rest of the magnetic
components. The resulting system describes the second-order
evolution of magnetic fields, which is frozen-in with the highly
conductive matter of a collapsing protogalaxy, within the limits
of the Zeldovich approximation. To obtain analytical solutions,
we recall that in the absence of rotation and acceleration,
the peculiar volume scalar is given by
\begin{equation}
\tilde{\vartheta}= {\lambda_1\over1+a\lambda_1}+
{\lambda_2\over1+a\lambda_2}+ {\lambda_3\over1+a\lambda_3}\,,
\label{vartheta}
\end{equation}
At the same time,
\begin{equation}
\tilde{\sigma}_{11}= {\lambda_1\over1+a\lambda_1}-
{1\over3}\,\vartheta\,,
\label{tsigma1}
\end{equation} while analogous expressions hold for the other
two shear eigenvalues. Note that $\lambda_1$, $\lambda_2$ and
$\lambda_3$ are the eigenvalues of the initial tidal field that
determine the nature of the collapse \citep{mata96,brun96}.
One-dimensional collapse along, say, the third eigen-direction
is characterized by $\lambda_1=0=\lambda_2$ and by $\lambda_3<0$.
In that case, the pancake singularity is reached as
$a\rightarrow-1/\lambda_3$. Spherically symmetric collapse,
on the other hand, has $\lambda_1=\lambda_2=\lambda_3=\lambda<0$.
Then, we have a point-like singularity when $a\rightarrow-1/\lambda$.

Substituting, Equations (\ref{vartheta}) and (\ref{tsigma1})
into the right-hand side of Equation (\ref{ZrB1}) we arrive at
the solution
\begin{equation}
B_1= B^0_1\left[{(1+a_0\lambda_2)(1+a_0\lambda_3)\over
(1+a\lambda_2)(1+a\lambda_3)}\right]\left({a_0\over a}\right)^2\,,
\label{ZB1}
\end{equation}
for the first of the magnetic components. A similar calculation
leads to exactly analogous equations for $B_2$ and $B_3$. The zero
suffix in the above indicates a given time during the protogalactic
collapse. The ratio $a_0/a$ in parentheses reflects
the magnetic dilution due to the background expansion and the
brackets monitor the increase in the field's strength caused
by the collapse of the protogalactic cloud. According to
(\ref{ZB1}), when dealing with pancake collapse along the
third eigen-direction, the $B_3$-component decays as $a^{-2}$,
while the other two increase arbitrarily as $a\rightarrow-1/\lambda_3$.
Alternatively, during a spherically symmetric contraction, the
magnetic field evolve as
\begin{equation}
B= B_0\left({1+a_0\lambda\over1+a\lambda}\right)^2
\left({a_0\over a}\right)^2\,.
\label{sphB}
\end{equation}
Here, all the magnetic components diverge as we approach the
point singularity (i.e. for $a\rightarrow-1/\lambda$). Comparing
the two results, we deduce that the anisotropic (pancake)
collapse leads to a stronger increase as long as $\lambda_3<\lambda$. 
The later is always satisfied, provided that the initial conditions 
are the same for both types of collapse, given that 
$\lambda_3=\tilde{\vartheta}_0/(1-a_0\tilde{\vartheta}_0)$ and 
$\lambda=\tilde{\vartheta}_0/(3-a_0\tilde{\vartheta}_0)$ 
[see expression (\ref{vartheta}) above].

Our qualitative analysis indicates that magnetic fields trapped
in an anisotropically contracting protogalactic cloud will
increase beyond the limits of the idealized spherically symmetric
collapse. Also, the amplified magnetic fields will end up
essentially confined to the galactic plane. Quantitatively,
the achieved final strength depends on the time the magnetic
back-reaction has grown strong enough to halt the collapse
\citep{zrs83}. Thus, the longer the anisotropic collapse persists,
the stronger the residual magnetic fields. The analytical
study of \citep{bmt03}, in particular, showed that
(realistically speaking) the anisotropy could add one or two
orders of magnitude to the magnetic strength achieved through
conventional isotropic compression. These results appear in
very good agreement with numerical studies simulating shear
and tidal effects on the magnetic field evolution in galaxies
and galaxy clusters \citep{rsb99,dbl02}.

\section{Summary}\label{s-summary}

The cosmic web, a network of filaments and nodes wherein most galaxies
reside, is a prediction of the highly successful $\Lambda$CDM
cosmology and appears to be borne out by by simulations and
observations.  The web is filled with an ionized plasma, the IGM,
which is expected to be permeated by magnetic fields.  In this
chapter, we reviewed recent developments in our theoretical
understanding of the nature and origin of IGMF with a special focus on
fields outside of clusters.  We addressed two basic questions: First,
``Can the process of structure formation generate seed magnetic fields
and amplify them?''  and second, ``What role do magnetic fields play
in structure formation and in other astrophysical processes?''  We
presented several plasma physics mechanisms for the generation of seed
magnetic fields, and showed that these fields could be amplified
during the first star formation and later during the formation of the
cosmic web.  We saw that magnetic fields with a strength of $\left< B
\right> \sim 10$ nG are expected in filaments at the present universe,
while magnetic fields should be stronger in and around
clusters/groups.  We then discussed the effects and implications of
magnetic fields on the formation of the first stars.  We also
presented a formal treatment of the evolution of density perturbations
in the presence of magnetic fields and their effects on the formation
of structures.

Magnetic fields in the IGM are difficult to be observed, mainly due to
the weak nature of them, as well as due to the diffuseness of the
media. However, the next generation radio interferometers including
the Square Kilometer Array (SKA), and upcoming SKA pathfinders, the
Australian SKA Pathfinder (ASKAP) and the South African Karoo Array
Telescope (MeerKAT), as well the Low Frequency Array (LOFAR) will
enable us to investigate magnetic fields outside clusters with
high-sensitivity observations of synchrotron radiation and RM
\citep[see, e.g., papers in][]{cr04}.

\begin{acknowledgements}

The work of DR was supported by the National Research Foundation
of Korea through grant 2007-0093860.
The work of DS was supported by a funding from the European
Community's Seventh Framework Programme (/FP7/2007-2013/) under
grant agreement No 229517.
The work of LMW was supported by a Discovery Grant from the
Natural Sciences and Engineering Research Council of Canada.

\end{acknowledgements}


\begin{thebibliography}{}

\bibitem[Abel et al.(2002)]{Abel02}
T. Abel, G. L. Bryan, M. L. Norman,
The formation of the first star in the universe.
Science {\bf 295}, 93-98 (2002). doi:10.1126/science.1063991

\bibitem[Achterberg and Wiersma(2007)]{achterberg:2007}
A. Achterberg, J. Wiersma,
The Weibel instability in relativistic plasmas. I. Linear theory.
\aap {\bf  475}, 1-36 (2007). doi:10.1051/0004-6361:20065365

\bibitem[Adams et al.(1996)]{adgr96}
J. Adams, U. H. Danielson, D. Grasso, H. R. Rubinstein,
Distortion of the acoustic peaks in the CMBR due to a primordial
magnetic field.
Phys. Lett. B {\bf 388}, 253-258 (1996).
doi:10.1016/S0370-2693(96)01171-9

\bibitem[Akahori and Ryu(2010)]{ar10}
T. Akahori, D. Ryu,
Faraday rotation measure due to the intergalactic magnetic field.
\apj {\bf 723}, 467-481 (2010). doi:10.1088/0004-637X/723/1/476

\bibitem[Aleksi$\acute{\rm c}$ et al.(2010)]{alek10}
J. Aleksi$\acute{\rm c}$ and MAGIC Collaboration,
Search for an extended VHE $\gamma$-ray emission from Mrk 421 and Mrk
501 with the MAGIC Telescope.
\aap {\bf 524}, id. A77 (2010). doi:10.1051/0004-6361/201014747


\bibitem[Arshakian et al.(2009)]{Arshakian09}
T. G. Arshakian, R. Beck, M. Krause, D. Sokoloff,
Evolution of magnetic fields in galaxies and future observational tests
with the Square Kilometre Array.
\aap {\bf 494}, 21-32 (2009). doi:10.1051/0004-6361:200810964

\bibitem[Barrow et al.(1997)]{bfs97}
J. D. Barrow, P. G. Ferreira, J. Silk,
Constraints on a primordial magnetic field.
\prl {\bf 78}, 3610-3613 (1997). doi:10.1103/PhysRevLett.78.3610

\bibitem[Barrow et al.(2007)]{bmt07}
J. D. Barrow, R. Maartens, C. G. Tsagas,
Cosmology with inhomogeneous magnetic fields.
\physrep {\bf 449}, 131-171 (2007). doi:10.1016/j.physrep.2007.04.006

\bibitem[Battaner et al.(1997)]{bfj97}
E. Battaner, E. Florido, J. Jimenez-Vicente,
Magnetic fields and large scale structure in a hot universe.
I. General equations.
\aap {\bf 326}, 13-22 (1997).

\bibitem[Bell(2004)]{bell04}
A. R. Bell,
Turbulent amplification of magnetic field and diffusive shock
acceleration of cosmic rays.
\mnras {\bf 353}, 550-558 (2004). doi:10.1111/j.1365-2966.2004.08097.x

\bibitem[Biermann(1950)]{bier50}
L. Biermann,
\"Uber den ursprung der magnetfelder auf sternen und im interstellaren
raum.
Z. Naturforsch. A {\bf 5}, 65-71 (1950).

\bibitem[Bonafede et al.(2010)]{bfmg10}
A. Bonafede, L. Feretti, M. Murgia, F. Govoni, G. Giovannini, G.,
D. Dallacasa, K. Dolag, G. B. Taylor,
The Coma cluster magnetic field from Faraday rotation measures.
\aap {\bf 513}, A30 (2010). doi:10.1051/0004-6361/200913696

\bibitem[Bond et al.(1996)]{bkp96}
J. R. Bond, L. Kofman, D. Pogosyan,
How filaments of galaxies are woven into the cosmic web.
Nature {\bf 380}, 603-606 (1996). doi:10.1038/380603a0

\bibitem[Brandenburg and Subramanian(2005)]{Brandenburg05}
A. Brandenburg, K. Subramanian,
Astrophysical magnetic fields and nonlinear dynamo theory.
\physrep {\bf 417}, 1-209 (2005). doi:10.1016/j.physrep.2005.06.005

\bibitem[Brandenburg et al.(2010)]{bss10}
A. Brandenburg, K. Subramanian, D. D. Sokoloff
Turbulent dynamos.
in this volume (2010).

\bibitem[Bromm and Loeb(2003)]{Bromm03}
V. Bromm, A. Loeb,
Formation of the first supermassive black holes.
\apj {\bf 596}, 34-46 (2003). doi:10.1086/377529

\bibitem[Brown et al.(2010)]{bfr10}
S. Brown, D. Farnsworth, L. Rudnick,
Cross-correlation of diffuse synchrotron and large-scale structures.
\mnras {\bf 402}, 2-6 (2010). doi:10.1111/j.1365-2966.2009.15867.x

\bibitem[Br\"uggen et al.(2010)]{bbrp10}
M. Br\"uggen, A. Bykov, D. Ryu, H. R\"ottgering,
Magnetic fields, relativistic particles, and shock waves in
cluster outskirts
in this volume (2010).

\bibitem[Bruni(1996)]{brun96}
M. Bruni,
Cosmological collapses of irrotational dust.
in {\it Mapping, measuring and modelling the universe},
Ed. P. Coles, V. Martinez, M.-J. Pons-Borderia,
ASP Conference Series {\bf 94}, 31-36 (1996).

\bibitem[Bruni et al.(2003)]{bmt03}
M. Bruni, R. Maartens, C. G. Tsagas,
Magnetic field amplification in cold dark matter anisotropic collapse.
\mnras {\bf 338}, 785-789 (2003). doi:10.1046/j.1365-8711.2003.06095.x

\bibitem[Caprini and Durrer(2002)]{cd02}
C. Caprini, R. Durrer,
Gravitational wave production: A strong constraint on primordial
magnetic fields.
\prd {\bf 65}, 023517 (2002). doi:10.1103/PhysRevD.65.023517

\bibitem[Carilli and Rawlings(2004)]{cr04}
C. L. Carilli, S. Rawlings,
{\it Science with the Square Kilometre Array}
(New A Rev. {\bf 48}, 2004).

\bibitem[Carilli and Taylor(2002)]{ct02}
C. L. Carilli, G. B. Taylor,
Cluster magnetic fields.
\araa {\bf 40}, 319-348 (2002).
doi:10.1146/annurev.astro.40.060401.093852

\bibitem[Cassano et al.(2008)]{cbvs08}
R. Cassano, G. Brunetti, T. Venturi, G. Setti, D. Dallacasa,
S. Giacintucci, S. Bardelli,
Revised statistics of radio halos and the reacceleration model.
\aap {\bf 480}, 687-697 (2008). doi:10.1051/0004-6361:20078986

\bibitem[Cassidy and Elford(1985)]{Cassidy85}
R. A. Cassidy, M. T. Elford,
The mobility of Li$^+$ ions in helium and argon.
Aust. J. Phys. {\bf 38}, 587-601 (1985).

\bibitem[Cen and Ostriker(1999)]{co99}
R. Cen, J. P. Ostriker,
Where are the baryons?
\apj {\bf 514}, 1-6 (1999). doi:10.1086/306949

\bibitem[Cen and Ostriker(2006)]{co06}
R. Cen, J. P. Ostriker,
Where are the baryons? II. Feedback effects.
\apj {\bf 650}, 560-572 (2006). doi:10.1086/506505

\bibitem[Cen et al.(2003)]{copw03}
R. Cen, J. P. Ostriker, J. X. Prochaska, A. M. Wolfe,
Metallicity evolution of damped Lyα systems in $\Lambda$CDM cosmology.
\apj {\bf 598}, 741-755 (2003). doi:10.1086/378881

\bibitem[Cho and Ryu(2009)]{cr09}
J. Cho, D. Ryu,
Characteristic lengths of magnetic field in magnetohydrodynamic
turbulence.
\apj {\bf 705}, L90-L94 (2009). doi:10.1088/0004-637X/705/1/L90

\bibitem[Cho and Vishniac(2000)]{cv00}
J. Cho, E. T. Vishniac,
The generation of magnetic fields through driven turbulence.
\apj {\bf 538}, 217-225 (2000). doi:10.1086/309127

\bibitem[Cho et al.(2009)]{cvbl09}
J. Cho, E. T. Vishniac, A. Beresnyak, A. Lazarian, D. Ryu,
Growth of magnetic fields induced by turbulent motions.
\apj {\bf 693}, 1449-1461 (2009). doi:10.1088/0004-637X/693/2/1449

\bibitem[Clark et al.(2011)]{Clark10}
P. C. Clark, S. C. O. Glover, R. S. Klessen, V. Bromm,
Gravitational fragmentation in turbulent primordial gas and the initial
mass function of Population III stars.
\apj {\bf 727}, id. 110 (2011). doi:10.1088/0004-637X/727/2/110

\bibitem[Clarke(2004)]{cla04}
T. E. Clarke,
Faraday rotation observations of magnetic fields in galaxy clusters.
J. Korean Astron. Soc. {\bf 37}, 337-342 (2004).

\bibitem[Clarke et al.(2001)]{ckb01}
T. E. Clarke, P. P. Kronberg, H. B\"ohringer,
A new radio-X-ray probe of galaxy cluster magnetic fields.
\apj {\bf 547}, L111-L114 (2001). doi:10.1086/318896

\bibitem[Das et al.(2008)]{dkrc08}
S. Das, H. Kang, D. Ryu, J. Cho,
Propagation of ultra-high-energy protons through the magnetized
cosmic web.
\apj {\bf 682}, 29-38 (2008). doi:10.1086/588278

\bibitem[Davies and Widrow(2000)]{dw00}
G. Davies, L.M. Widrow,
A possible mechanism for generating galactic magnetic fields.
\apj {\bf 540}, 755-764 (2000). doi:10.1086/309358

\bibitem[Dermer et al.(2011)]{dermer}
C. D. Dermer, M. Cavadini, S. Razzaque, J. D. Finke, J. Chiang, B. Lott,
Time delay of cascade radiation for TeV blazars and the measurement of
the intergalactic magnetic field.
\apjl {\bf 733}, L21-L24 (2011). doi:10.1088/2041-8205/733/2/L21

\bibitem[Dickinson et al.(1982)]{Dickinson82}
A. S. Dickinson, M. S. Lee, W. A. Lester, Jr.,
Close-coupling calculation of Li$^+ -$H$_2$ diffusion cross sections.
J. Phys. B {\bf 15}, 1371-1376 (1982). doi:10.1088/0022-3700/15/9/013

\bibitem[Dolag et al.(2002)]{dbl02}
K. Dolag, M. Bartelmann, H. Lesch,
Evolution and structure of magnetic fields in simulated galaxy clusters.
\aap {\bf 387}, 383-395 (2002). doi:10.1051/0004-6361:20020241


\bibitem[Donnert et al.(2009)]{ddlm09}
J. Donnert, K. Dolag, H. Lesch, E. M\"{u}ller,
Cluster magnetic fields from galactic outflows.
\mnras {\bf 392}, 1008-1021 (2009). doi:10.1111/j.1365-2966.2008.14132.x

\bibitem[Fennelly and Evans(1980)]{fe80}
A. J. Fennelly, C. R. Evans,
Magnetohydrodynamic perturbations of Robertson-Walker universes and of
anisotropic Bianchi type-I universes.
Nuovo Cim. B {\bf 60}, 1-45 (1980). doi:10.1007/BF02723065

\bibitem[Federrath et al.(2011)]{fssb11}
C. Federrath, S. Sur, D. R. G. Schleicher, R. Banerjee, R. S. Klessen,
A new jeans resolution criterion for (M)HD simulations of self-gravitating gas:
Application to magnetic field amplification by gravity-driven turbulence.
\apj {\bf 71}, id. 62 (2011). doi:10.1088/0004-637X/731/1/62

\bibitem[Fried(1959)]{fried:1959}
B. D. Fried,
Mechanism for instability of transverse plasma waves.
Phys. Fluids {\bf 2}, 337 (1959). doi:10.1063/1.1705933

\bibitem[Fromang et al.(2004)]{Fromang04}
S. Fromang, S. A. Balbus, C. Terquem, J. De Villiers,
Evolution of self-gravitating magnetized disks. II. Interaction between
magnetohydrodynamic turbulence and gravitational instabilities.
\apj {\bf 616}, 364-375 (2004). doi:10.1086/424829


\bibitem[Glover and Savin(2009)]{Glover09}
S. C. O. Glover, D. W. Savin,
Is H$_+^3$ cooling ever important in primordial gas?
\mnras {\bf 393}, 911-948 (2009). doi:10.1111/j.1365-2966.2008.14156.x


\bibitem[Govoni and Feretti(2004)]{gf04}
F. Govoni, L. Feretti,
Magnetic fields in clusters of galaxies.
Int. J. Mod. Phys. D {\bf 13}, 1549-1594 (2004).
doi:10.1142/S0218271804005080

\bibitem[Grasso and Rubinstein(2001)]{Grasso01}
D. Grasso, H. R. Rubinstein,
Magnetic fields in the early universe.
\physrep {\bf 348}, 163-266 (2001). doi:10.1016/S0370-1573(00)00110-1

\bibitem[Gruzinov(2001)]{gruzinov:2001}
A. Gruzinov,
Gamma-ray burst phenomenology, shock dynamics, and the first magnetic
fields.
\apjl {\bf 563}, L15-L18 (2001). doi:10.1086/324223

\bibitem[Gruzinov and Waxman(1999)]{gruzinov:1999}
A. Gruzinov, E. Waxman,
Gamma-ray burst afterglow: Polarization and analytic light curves.
\apj {\bf 511}, 852-861 (1999).doi:10.1086/306720

\bibitem[Guidetti et al.(2008)]{gmgp08}
D. Guidetti, M. Murgia, F. Govoni, P. Parma, L. Gregorini,
H. R. deRuiter, R. A. Cameron, R. Fanti,
The intracluster magnetic field power spectrum in Abell 2382.
\aap {\bf 483}, 699-713 (2008). doi:10.1051/0004-6361:20078576

\bibitem[Haugen et al.(2004a)]{Haugen04a}
N. E. L. Haugen, A. Brandenburg, W. Dobler,
Simulations of nonhelical hydromagnetic turbulence.
\pre {\bf 70}, 016308 (2004a). doi:10.1103/PhysRevE.70.016308

\bibitem[Haugen et al.(2004b)]{Haugen04b}
N. E. L. Haugen, A. Brandenburg, W. Dobler,
High-resolution simulations of nonhelical MHD turbulence.
Astrophys. Space Sci. {\bf 292}, 53-60 (2004b).
doi:10.1023/B:ASTR.0000045000.08395.a3

\bibitem[Haugen et al.(2004c)]{Haugen04c}
N. E. L. Haugen, A. Brandenburg, A. J. Mee,
Mach number dependence of the onset of dynamo action.
\mnras {\bf 353}, 947-952 (2004c). doi:10.1111/j.1365-2966.2004.08127.x

\bibitem[Hoyle(1969)]{hoyle69}
F. Hoyle,
Magnetic fields and highly condensed objects.
Nature {\bf 223}, 936 (1969). doi:10.1038/223936a0

\bibitem[Jaroschek and Hoshino(2009)]{jaroschek:2009}
C. H. Jaroschek, M. Hoshino,
Radiation dominated relativistic current sheets.
\prl {\bf 103}, 075002 (2009). doi:10.1103/PhysRevLett.103.075002

\bibitem[Jaroschek et al.(2008)]{jaroschek:2008}
C. H. Jaroschek, M. Hoshino, H. Lesch, R. A. Treumann,
Stochastic particle acceleration by the forced interaction of
relativistic current sheets.
Adv. Space Res. {\bf 41}, 481-490 (2008). doi:10.1016/j.asr.2007.07.001

\bibitem[Jaroschek et al.(2005)]{jaroschek:2005}
C. H. Jaroschek, H. Lesch, R. A. Treumann,
Ultra-relativistic plasma shell collision s in gamma-ray burst sources:
Dimensional effects and the final steady state magnetic field.
\apj {\bf 618}, 822-831 (2005). doi:10.1086/426066

\bibitem[Kang et al.(2007)]{krco07}
H. Kang, D. Ryu, R. Cen, J. P. Ostriker,
Cosmological shock waves in the large-scale structure of the universe:
Nongravitational effects.
\apj {\bf 669}, 729-740 (2007). doi:10.1086/521717

\bibitem[Kang et al.(2005)]{krcs05}
H. Kang, D. Ryu, R. Cen, D. Song,
Shock-heated gas in the large-scale structure of the universe.
\apj {\bf 620}, 21-30 (2005). doi:10.1086/426931


\bibitem[Kim et al.(1990)]{kkdl90}
K. T. Kim, P. P. Kronberg, P. D. Dewdney, T. L. Landecker,
The halo and magnetic field of the Coma cluster of galaxies.
\apj {\bf 355}, 29-37 (1990). doi:10.1086/168737

\bibitem[Kim et al.(1989)]{kkgt89}
K. T. Kim, P. P. Kronberg, G. Giovannini, T. Venturi,
Discovery of intergalactic radio emission in the Coma-A1367
supercluster.
Nature {\bf 341}, 720-723 (1989). 10.1038/341720a0

\bibitem[Kim et al.(1996)]{kor96}
E.-J. Kim, A. V. Olinto, R. Rosner,
Generation of density perturbations by primordial magnetic fields.
\apj {\bf 468}, 28-50 (1996). doi:10.1086/177667

\bibitem[Kolb and Turner(1990)]{Kolb90}
E. W. Kolb, M. S. Turner,
{\it The early universe}
(Addison-Wesley, Redwood City, 1990).

\bibitem[Krause et al.(2009)]{kra09}
M. Krause, P. Alexander, R. Bolton, J. Geisb\"{u}sch, D. A. Green, J. Riley,
Measurements of the cosmological evolution of magnetic fields with the
Square Kilometre Array.
\mnras {\bf 400}, 646-656 (2009). doi:10.1111/j.1365-2966.2009.15489.x 

\bibitem[Kronberg et al.(2001)]{kdlc01}
P. P. Kronberg, Q. W. Dufton, H. Li, S. A. Colgate,
Magnetic energy of the intergalactic medium from galactic black holes.
\apj {\bf 560}, 178-186 (2001). doi:10.1086/322767

\bibitem[Kulsrud et al.(1997)]{kcor97}
R. M. Kulsrud, R. Cen, J. P. Ostriker, D. Ryu,
The protogalactic origin for cosmic magnetic fields.
\apj {\bf 480}, 481-491 (1997). doi:10.1086/303987

\bibitem[Kulsrud and Zweibel(2008)]{kz07}
R. M. Kulsrud, E. G. Zweibel,
On the origin of cosmic magnetic fields.
Rep. Prog. Phys. {\bf 71}, 046901 (2008).
doi:10.1088/0034-4885/71/4/046901

\bibitem[Large et al.(1959)]{lmh59}
M. I. Large, D. S. Mathewson, C. G. T. Haslam,
A high-resolution survey of the Coma cluster of galaxies at 408 Mc./s.
Nature {\bf 183}, 1663-1664 (1959). doi:10.1038/1831663a0

\bibitem[Larson(1969)]{Larson69}
R. B. Larson,
Numerical calculations of the dynamics of collapsing proto-star.
\mnras {\bf 145}, 271-295 (1969).

\bibitem[Lazarian(1992)]{lazarian:1992}
A. Lazarian,
Diffusion-generated electromotive force and seed magnetic field problem.
\aap {\bf 264}, 326-330 (1992).

\bibitem[Machida et al.(2008)]{Machida08}
M. N. Machida, S. Inutsuka, T. Matsumoto, T.
High- and low-velocity magnetized outflows in the star formation
process in a gravitationally collapsing cloud.
\apj {\bf 676}, 1088-1108 (2008). doi:10.1086/528364

\bibitem[Machida et al.(2006)]{Machida06}
M. N. Machida, K. Omukai, T. Matsumoto, S. Inutsuka,
The first jets in the universe: protostellar jets from the first
stars.
\apjl {\bf 647}, L1-L4 (2006). doi:10.1086/507326

\bibitem[Maki and Susa(2004)]{Maki04}
H. Maki, H. Susa,
Dissipation of magnetic flux in primordial gas clouds.
\apj {\bf 609}, 467-473 (2004). doi:10.1086/421103

\bibitem[Matarrese(1996)]{mata96}
S. Matarrese,
Relativistic Cosmology: from Superhorizon to Small Scales.
in {\it Dark matter in the universe},
Ed. S. Bonometto, J. R. Primack, A. Provenzale,
(IOS Press, Oxford, 1996) pp. 601-628.

\bibitem[Medvedev and Loeb(1999)]{medvedev:1999}
M. V. Medvedev, A. Loeb,
Generation of Magnetic Fields in the Relativistic Shock of Gamma-Ray
Burst Sources.
\apj {\bf 526}, 697-706 (1999). doi:10.1086/308038

\bibitem[Medvedev et al.(2006)]{msk06}
M. V. Medvedev, L. O. Silva, M. Kamionkowski,
Cluster magnetic fields from large-scale structure and galaxy
cluster shocks.
\apjl {\bf 642}, L1-L4 (2006). doi:10.1086/504470

\bibitem[Miniati and Bell(2011)]{mb10}
F. Miniati, A. R. Bell,
Resistive magnetic field generation at cosmic dawn.
\apj {\bf 729}, id. 73 (2011). doi:10.1088/0004-637X/729/1/73

\bibitem[Myers and Khersonsky(1995)]{Myers95}
P. C. Myers, V. K. Khersonsky,
On magnetic turbulence in interstellar clouds.
\apj {\bf 442}, 186-196 (1995). doi:10.1086/175434

\bibitem[Nagai et al.(2007)]{nvk07}
D. Nagai, A. Vikhlinin, A. V. Kravtsov,
Testing X-ray measurements of galaxy clusters with cosmological
simulations.
\apj {\bf 655}, 98-108 (2007). doi:10.1086/509868

\bibitem[Neronov and Vovk(2010)]{nv10}
A. Neronov, I. Vovk
Evidence for strong extragalactic magnetic fields from Fermi
observations of TeV Blazars.
Science {\bf 328}, 73-75 (2010). doi:10.1126/science.1184192

\bibitem[Nishikawa et al.(2009)]{nishikawa:2009}
K. I. Nishikawa, J. Niemiec, P. E. Hardee, M. Medvedev, H. Sol,
Y. Mizuno, B. Zhang, M. Pohl, M. Oka, D. H. Hartmann,
Weibel instability and associated strong fields in a fully
three-dimensional simulation of a relativistic shock.
\apjl {\bf 698}, L10-L13 (2009). doi:10.1088/0004-637X/698/1/L10

\bibitem[Peebles(1968)]{Peebles68}
P. J. E. Peebles,
Recombination of the primeval plasma.
\apj {\bf 153}, 1-11 (1968). doi:10.1086/149628

\bibitem[Penston(1969)]{Penston69}
M. V. Penston,
Dynamics of self-gravitating gaseous spheres-III. Analytical results
in the free-fall of isothermal cases.
\mnras {\bf 144}, 425-448 (1969).

\bibitem[Pfrommer et al.(2006)]{psej06}
C. Pfrommer, V. Springel, T. A. En{\ss}lin, M. Jubelgas,
Detecting shock waves in cosmological smoothed particle hydrodynamics
simulations.
\mnras {\bf 367}, 113-131 (2006). doi:10.1111/j.1365-2966.2005.09953.x

\bibitem[Pinto and Galli(2008)]{Pinto08b}
C. Pinto, D. Galli,
Three-fluid plasmas in star formation. II. Momentum transfer rate
coefficients.
\aap {\bf 484}, 17-28 (2008). doi:10.1051/0004-6361:20078819

\bibitem[Pinto et al.(2008)]{Pinto08a}
C. Pinto, D. Galli, F. Bacciotti,
Three-fluid plasmas in star formation. I. Magneto-hydrodynamic
equations.
\aap {\bf 484}, 1-15 (2008). doi:10.1051/0004-6361:20078818

\bibitem[Plaga(1995)]{plag95}
R. Plaga,
Detecting intergalactic magnetic fields using time delays in pulses
of $\gamma$-rays.
Nature {\bf 374}, 430-432 (1995). doi:10.1038/374430a0

\bibitem[Pudritz and Silk(1989)]{Pudritz89}
R. E. Pudritz, J. Silk,
The origin of magnetic fields and primordial stars in protogalaxies.
\apj {\bf 342}, 650-659 (1989). doi:10.1086/167625

\bibitem[R{\o}eggen et al.(2002)]{Roeggen02}
I. R{\o}eggen, H. R. Skullerud, T. H. L{\o}vaas, D. K. Dysthe,
The Li$^+ -$H$_2$ system in a rigid-rotor approximation: potential
energy surface and transport coefficients.
J. Phys. B {\bf 35}, 1707-1725 (2002). doi:10.1088/0953-4075/35/7/309

\bibitem[Roettiger et al.(1999)]{rsb99}
K. Roettiger, J. M. Stone, J.O. Burns,
Magnetic field evolution in merging clusters of galaxies.
\apj {\bf 518}, 594-602 (1999). doi:10.1086/307298

\bibitem[Ryu et al.(2010)]{rdk10}
D. Ryu, S. Das, H. Kang,
Intergalactic magnetic field and arrival direction of
ultra-high-energy Protons.
\apj {\bf 710}, 1422-1431 (2010). doi:10.1088/0004-637X/710/2/1422

\bibitem[Ryu and Kang(2008)]{rk08}
D. Ryu, H. Kang,
Vorticity and turbulence in the large-scale structure of the universe.
in {\it Numerical Modeling of Space Plasma Flows: Astronum 2007},
Ed. N. V. Pogorelov, E. Audit, G. P. Zank,
ASP Conference Series {\bf 385}, 44-49 (2008).

\bibitem[Ryu et al.(1998)]{rkb98}
D. Ryu, H. Kang, P. L. Biermann,
Cosmic magnetic fields in large scale filaments and sheets.
\aap {\bf 335}, 19-25 (1998).

\bibitem[Ryu et al.(2008)]{rkcd08}
D. Ryu, H. Kang, J. Cho, S. Das,
Turbulence and magnetic fields in the large-scale structure of
the universe.
Science {\bf 320}, 909-912 (2008). doi:10.1126/science.1154923

\bibitem[Ryu et al.(2003)]{rkhj03}
D. Ryu, H. Kang, E. Hallman, T. W. Jones,
Cosmological shock waves and their role in the large-scale structure
of the universe.
\apj {\bf 593}, 599-610 (2003). doi:10.1086/376723

\bibitem[Ryu et al.(1993)]{rokc93}
D. Ryu, J. P. Ostriker, H. Kang, R. Cen,
A cosmological hydrodynamic code based on the total variation
diminishing scheme.
\apj {\bf 414}, 1-19 (1993). doi:10.1086/173051

\bibitem[Ruzmaikina and Ruzmaikin(1971)]{rr71}
T. V. Ruzmaikina, A. A. Ruzmaikin,
Gravitational stability of an expanding universe in the presence
of a magnetic field.
Sov. Astron. {\bf 14}, 963-966 (1971).

\bibitem[Sakai et al.(2004)]{sakai:2004}
J. I. Sakai, R. Schlickeiser,  P. K. Shukla,
Simulation studies of magnetic field generation in cosmological
plasmas.
Phys. Lett. A {\bf 330}, 384-389 (1999).
doi:10.1016/j.physleta.2004.08.007

\bibitem[Schekochihin et al.(2004)]{Schekochihin04}
A. A. Schekochihin, S. C. Cowley, S. F. Taylor, J. L. Maron,
J. C. McWilliams,
Simulations of the small-scale turbulent dynamo.
\apj {\bf 612}, 276-307 (2004). doi:10.1086/422547

\bibitem[Schekochihin et al.(2010)]{sbfk10}
A. A. Schekochihin, M. Br\"uggen, L. Feretti, M. W. Kunz, L. Rudnick,
Magnetic fields in galaxy clusters: why bother?
in this volume (2010).

\bibitem[Schleicher et al.(2008)]{Schleicher08b}
D. R. G. Schleicher, R. Banerjee, R. S. Klessen,
Reionization: A probe for the stellar population and the physics of
the early universe.
\prd {\bf 78}, 083005 (2008). doi:10.1103/PhysRevD.78.083005

\bibitem[Schleicher et al.(2010)]{Schleicher10c}
D. R. G. Schleicher, R. Banerjee, S. Sur, T. G. Arshakian,
R. S. Klessen, R. Beck, M. Spaans,
Small-scale dynamo action during the formation of the first stars
and galaxies. I. The ideal MHD limit.
\aap, {\bf 522}, id. A115 (2010). doi:10.1051/0004-6361/201015184

\bibitem[Schleicher et al.(2009)]{Schleicher09prim}
D. R. G. Schleicher, D. Galli, S. C. O. Glover, R. Banerjee,
F. Palla, R. Schneider, R. S. Klessen,
The influence of magnetic fields on the thermodynamics of primordial
star formation.
\apj {\bf 703}, 1096-1106 (2009). doi:10.1088/0004-637X/703/1/1096

\bibitem[Schlickeiser and Shukla(2003)]{ss03}
R. Schlickeiser, P. K. Shukla,
Cosmological magnetic field generation by the Weibel instability.
\apjl {\bf 599}, L57-L60 (2003). doi:10.1086/381246

\bibitem[Schuecker et al.(2004)]{sfmb04}
P. Schuecker, A. Finoguenov, F. Miniati, H. B\"ohringer, H.,
U. G. Briel,
Probing turbulence in the Coma galaxy cluster.
\aap {\bf 426}, 387-397 (2004). doi:10.1051/0004-6361:20041039

\bibitem[Seager et al.(1999)]{Seager99}
S. Seager, D. D. Sasselov, D. Scott,
A new calculation of the recombination epoch.
\apjl {\bf 523}, L1-L5 (1999). doi:10.1086/312250

\bibitem[Sethi et al.(2008)]{Sethi08}
S. K. Sethi, B. B. Nath, K. Subramanian,
Primordial magnetic fields and formation of molecular hydrogen.
\mnras {\bf 387}, 1589-1596 (2008).
doi:10.1111/j.1365-2966.2008.13302.x

\bibitem[Sethi and Subramanian(2005)]{ss05}
S. K. Sethi, K. Subramanian,
Primordial magnetic fields in the post-recombination era and
early reionization.
\mnras {\bf 356}, 778-788 (2005). doi:10.1111/j.1365-2966.2004.08520.x

\bibitem[Silk and Langer(2006)]{Silk06}
J. Silk, M. Langer,
On the first generation of stars.
\mnras {\bf 371}, 444-450 (2006). doi:10.1111/j.1365-2966.2006.10689.x

\bibitem[Sitenko(1967)]{sitenko:1967}
A. G. Sitenko,
{\it Electromagnetic fluctuations in plasma}, Chap. 4
(Academic Press, New York, 1967).

\bibitem[Skillman et al.(2008)]{sohb08}
S. W. Skillman, B. W. O'Shea, E. Hallman, J. O. Burns,
M. L. Norman,
Cosmological shocks in adaptive mesh refinement simulations and
the acceleration of cosmic rays.
\apj {\bf 689}, 1063-1077 (2008). doi:10.1086/592496

\bibitem[de Souza and Opher(2010)]{deSouza10}
R. S. de Souza, R. Opher,
Origin of magnetic fields in galaxies.
\prd {\bf 81}, 067301 (2010). doi:10.1103/PhysRevD.81.067301

\bibitem[Subramanian(1999)]{Subramanian99}
K. Subramanian,
Unified treatment of small- and large-scale dynamos in helical
turbulence.
\prl {\bf 83}, 2957-2960 (1999). doi:10.1103/PhysRevLett.83.2957

\bibitem[Subramanian and Barrow(1998)]{sb98}
K. Subramanian, J. D. Barrow,
Magnetohydrodynamics in the early universe and the damping of
nonlinear Alfv\'en waves.
\prd {\bf 58}, 083502 (1998). doi:10.1103/PhysRevD.58.083502

\bibitem[Subramanian et al.(1994)]{subramanian:1994}
K. Subramanian, D. Narashimha, S. M. Chitre,
Thermal generation of cosmological seed magnetic fiels in
ionization fronts.
\mnras {\bf 271}, L15-L18 (1994).

\bibitem[Sur et al.(2010)]{ssbf10}
S. Sur, D. R. G. Schleicher, R. Banerjee, C. Federrath, R. S. Klessen,
The generation of strong magnetic fields during the formation of
the first stars.
\apjl {\bf 721}, L134-L138 (2010). doi:10.1088/2041-8205/721/2/L134

\bibitem[Syrovatskii(1970)]{syro70}
S. I. Syrovatskii,
in {\it Interstellar Gas Dynamics},
ed. H. J. Habing,
IAU Symposium No. 39, 192-192 (Springer-Verlag, New York, 1970)

\bibitem[Takami and Sato(2008)]{ts08}
H. Takami, K. Sato,
Distortion of ultra-high-energy sky by galactic magnetic field.
\apj {\bf 681}, 1279-1286 (2008). doi:10.1086/588513

\bibitem[Tan and Blackman(2004)]{Tan04}
J. C. Tan, E. G. Blackman,
Protostellar disk dynamos and hydromagnetic outflows in
primordial star formation.
\apj {\bf 603}, 401-413 (2004). doi:10.1086/381668

\bibitem[Tashiro and Sugiyama(2006)]{Tashiro06a}
H. Tashiro, N. Sugiyama,
Early reionization with primordial magnetic fields.
\mnras {\bf 368}, 965-970 (2006). doi:10.1111/j.1365-2966.2006.10178.x

\bibitem[The Pierre Auger Collaboration(2007)]{auger07a}
The Pierre Auger Collaboration,
Correlation of the highest-energy cosmic rays with nearby
extragalactic objects.
Science {\bf 318}, 938-943 (2007). doi:10.1126/science.1151124

\bibitem[Treumann(2010)]{treumann:2010}
R. A. Treumann, R. Nakamura, and W. Baumjohann,
Collisionless reconnection: Mechanism of self-ignition in thin plane
current homogeneous sheets.
Ann. Geophys. {\bf 28}, 1935-1943 (2010). doi:10.5194/anngeo-28-1935-2010

\bibitem[Tsagas(2002)]{ts02}
C. G. Tsagas,
Gravitational waves and cosmic magnetism: a cosmological approach.
Class. Quantum Grav. {\bf 19}, 3709-3722 (2002).
doi:10.1088/0264-9381/19/14/311

\bibitem[Tsagas and Barrow(1997)]{tb97}
C. G. Tsagas, J. D. Barrow,
A gauge-invariant analysis of magnetic fields in general-relativistic
cosmology.
Class. Quantum Grav. {\bf 14}, 2539-2562 (1997).
doi:10.1088/0264-9381/14/9/011

\bibitem[Tsagas and Maartens(2000a)]{tm00a}
C. G. Tsagas, R. Maartens,
Magnetized cosmological perturbations.
\prd {\bf 61}, 083519 (2000a). doi:10.1103/PhysRevD.61.083519

\bibitem[Tsagas and Maartens(2000b)]{tm00b}
C. G. Tsagas, R. Maartens,
Cosmological perturbations on a magnetized Bianchi I background.
Class. Quantum Grav. {\bf 17} 2215-2241 (2000b).
doi:10.1088/0264-9381/17/11/305

\bibitem[Turk et al.(2009)]{Turk10}
M. J. Turk, T. Abel, B. O'Shea,
The formation of population III binaries from cosmological
initial conditions.
Science {\bf 325}, 601-605 (2009). doi:10.1126/science.1173540

\bibitem[Vazza et al.(2009)]{vbg09}
F. Vazza, G. Brunetti, C. Gheller, C.,
Shock waves in Eulerian cosmological simulations: main properties
and acceleration of cosmic rays.
\mnras {\bf 395}, 1333-1354 (2009).
doi:10.1111/j.1365-2966.2009.14691.x

\bibitem[Vogt and En{\ss}lin(2005)]{ve05}
C. Vogt, T. En{\ss}lin,
A Bayesian view on Faraday rotation maps Seeing the magnetic power
spectra in galaxy clusters.
\aap {\bf 434}, 67-76 (2005). doi:10.1051/0004-6361:20041839

\bibitem[Wang(2010)]{wa10}
S. Wang,
New primordial-magnetic-field limit from the latest LIGO S5 data.
\prd {\bf 81}, 023002 (2010). doi:10.1103/PhysRevD.81.023002

\bibitem[Wasserman(1978)]{wa78}
I. Wasserman,
On the origins of galaxies, galactic angular momenta and galactic
magnetic fields.
\apj {\bf 224}, 337-343 (1978). doi:10.1086/156381

\bibitem[Weibel(1959)]{weibel:1959}
E. Weibel,
Spontaneously growing transverse  waves in a plasma due to
anisotropic velocity distribution.
\prl {\bf 2}, 83-84 (1959). doi:10.1103/PhysRevLett.2.83

\bibitem[Widrow(2002)]{widrow:2002}
L. M. Widrow,
Origin of galactic and extragalactic magnetic fields.
Rev. Mod. Phys. {\bf 74}, 775-823 (2002). doi:10.1103/RevModPhys.74.775

\bibitem[Widrow et al.(2010)]{wrss10}
L. M. Widrow, D. Ryu, D. R. G. Schleicher, K. Subramanian,
R. A. Treumann, C. Tsagas,
The first magnetic fields.
in this volume (2010).

\bibitem[Xu et al.(2008)]{xocn08}
H. Xu, B. W. O'Shea, D. C. Collins, M. L. Norman, H. Li, S. Li,
The Biermann battery in cosmological MHD simulations of population
III star formation.
\apjl {\bf 688}, L57-L60 (2008). doi:10.1086/595617

\bibitem[Xu et al.(2006)]{xkhd06}
Y. Xu, P. P. Kronberg, S. Habib, Q. W. Dufton,
A Faraday rotation search for magnetic fields in large-scale structure.
\apj {\bf 637}, 19-26 (2006). doi:10.1086/498336

\bibitem[Yoshida et al.(2008)]{Yoshida08}
N. Yoshida, K. Omukai, L. Hernquist,
Protostar formation in the early universe.
Science {\bf 321}, 669-671 (2008). doi:10.1126/science.1160259

\bibitem[Zeldovich(1970a)]{zeld70a}
Y. B. Zeldovich,
The Hypothesis of Cosmological Magnetic Inhomogeneity.
Sov. Astron. {\bf 13}, 608-611 (1970a).

\bibitem[Zeldovich(1970b)]{zeld70b}
Y. B. Zeldovich,
Separation of uniform matter into parts under the action of
gravitation.
Astrofizika {\bf 6}, 319-335 (1970b).

\bibitem[Zeldovich et al.(1969)]{Zeldovich69}
Y. B. Zeldovich, V. G. Kurt, R. A. Sunyaev,
Recombination of hydrogen in the hot model of the universe.
Sov. Phys. JETP {\bf 28}, 146-150 (1969).

\bibitem[Zeldovich et al.(1983)]{zrs83}
Y. B. Zeldovich, A. A. Ruzmaikina, D. D. Sokolov,
{\it Magnetic fields in astrophysics}
(Gordon and Breach Science Publishers, New York, 1983).

\end{thebibliography}
\end{document}